\documentclass{LMCS}

\usepackage{amsmath,amssymb,gastex,rotating,array,color,multirow}
\usepackage{hyperref}

\newlength{\cellsize}
\newcommand{\cell}[1]{
  \minipage[c][\cellsize][c]{\cellsize}%
  \centering{#1}\endminipage}

\newcommand{\stile}[4]{
  \begin{array}{|@{}c@{}|@{}c@{}|}
    \hline
    \cell{$#1$}
    &  \cell{$#2$}  
    \\\hline
    \cell{$#3$}
    &  \cell{$#4$}  
    \\\hline
  \end{array}}

\newcommand{\PBS}[1]{\let\temp=\\#1\let\\=\temp}

\newcommand{\era}[2]{\overset{#1}{\underset{#2}{\longrightarrow}}}
\newcommand{\erb}[1]{\overset{#1}{\longrightarrow}}

\newcommand{\eRb}[1]{\overset{#1}{\Longrightarrow}}

\newcommand{\td}{\text{\tiny $\#$}}
\newcommand{\sd}{\text{\small $\#$}}
\newcommand{\eps}{\varepsilon}
\newcommand{\fr}[1]{\mathrm{fr}\left(#1\right)}

\newcommand{\sep}{\;|\;}

\newcommand{\intro}[1]{\emph{#1}}

 \makeatletter 
 \renewcommand{\section}{\@startsection
   {section}{1}%
   \z@{1.3\linespacing\@plus\linespacing}%
   {.5\linespacing}%
   {\normalfont\scshape\centering}}
 \renewcommand{\subsection}{\@startsection
   {subsection}{2}%
   \z@{.5\linespacing\@plus.7\linespacing}%
   {.5\linespacing}%
   {\normalfont\bfseries}}
 \renewcommand{\subsubsection}{\@startsection
   {subsubsection}{3}%
   \z@{.5\linespacing\@plus.7\linespacing}%
   {.5\linespacing}%
   {\normalfont\itshape}}
 \makeatother

\newcommand{\automate}{A}

\def\doi{2 (2:6) 2006}
\lmcsheading%
{\doi}
{24}
{}
{}
{Jan.~31, 2005}
{Jul.~19, 2006}
{}

\begin{document}

\title{Context-Sensitive Languages, Rational Graphs and Determinism}

\author[A.\ Carayol]{Arnaud Carayol\rsuper a} 
\address{{\lsuper a}{\sc Irisa} -- Campus de Beaulieu -- 35042 Rennes Cedex --
  France}
\email{Arnaud.Carayol@irisa.fr}

\author[A.\ Meyer]{Antoine Meyer\rsuper b} 
\address{{\lsuper b}{\sc Liafa} -- Universit\'e de Paris 7 -- 2
  place Jussieu, case 7014, 75251 Paris Cedex 05 -- France}
\email{Antoine.Meyer@liafa.jussieu.fr}

\keywords{language theory, infinite graphs, automata, determinism}
\subjclass{F.4.1}

\begin{abstract}
  We investigate families of infinite automata for context-sensitive
  languages. An infinite automaton is an infinite labeled graph with
  two sets of initial and final vertices. Its language is the set of
  all words labelling a path from an initial vertex to a final vertex.
  In 2001, Morvan and Stirling proved that rational graphs accept the
  context-sensitive languages between rational sets of initial and
  final vertices. This result was later extended to sub-families of
  rational graphs defined by more restricted classes of transducers.

  Our contribution is to provide syntactical and self-contained proofs
  of the above results, when earlier constructions relied on a
  non-trivial normal form of context-sensitive grammars defined by
  Penttonen in the 1970's. These new proof techniques enable us to
  summarize and refine these results by considering several
  sub-families defined by restrictions on the type of transducers, the
  degree of the graph or the size of the set of initial vertices.
\end{abstract}

\maketitle

\section{Introduction}
\label{sect:intro}

One of the cornerstones of formal language theory is the well-known
hierarchy introduced by Chomsky in \cite{Chomsky59}. It consists of
the regular, context-free, context-sensitive and recursively
enumerable languages. This hierarchy was originally defined by
imposing syntactical restrictions on the rules of grammars generating
the languages. These four families of languages as well as some of
their sub-families have been extensively studied. In particular, they
were given alternative characterizations in terms of finite acceptors.
They are respectively accepted by finite automata, pushdown automata,
linearly bounded automata and Turing machines.  Recently, these
families of languages have been characterised by families of infinite
automata. An infinite automaton is a labelled countable graph together
with a set of initial and a set of final vertices.  The language it
accepts (or simply its \emph{language}) is the set of all words
labelling a path from an initial vertex to a final vertex.  In
\cite{Caucal02}, a summary of four families of graphs corresponding to
the four families in the Chomsky hierarchy was given: they are
respectively the finite graphs, \intro{prefix-recognisable} graphs
\cite{Caucal96,Caucal03pr}, \intro{rational} graphs \cite{Morvan00}
and transition graphs of Turing machines \cite{Caucal03tm} (for a
survey, see for instance \cite{Thomas01}).

This work specifically deals with a family of infinite automata for
context-sensitive languages. The first result on this topic is due to
Morvan and Stirling \cite{Morvan01a}, who showed that the languages
accepted by rational graphs, whose vertices are words and whose edges
are defined by rational transducers, taken between rational or finite
sets of vertices, are precisely the context-sensitive languages. This
result was later extended by Rispal \cite{Rispal02} to the more
restricted families of \intro{synchronized} rational graphs, and even
to \intro{synchronous} graphs. A summary can be found in
\cite{Morvan04}.  All proofs provided in these works use
context-sensitive grammars in Penttonen normal form \cite{Penttonen74}
to characterize context-sensitive languages, which has two main
drawbacks. First, this normal form is far from being obvious, and the
proofs and constructions provided in \cite{Penttonen74} are known to
be difficult. Second, and more importantly, there is no grammar-based
characterization of deterministic context-sensitive languages, which
forbids one to adapt these results to the deterministic case.

Our main contributions are a new syntactical proof of the theorem by
Stirling and Morvan based on the thight correspondance between tiling
systems and synchronized graphs and an in depth study of the trade off
between the structure of the rational graphs (number of initial
vertices and out-degree), the transducers defining them, and the
family of languages they accept, as summarized in Table
\ref{tab:classif}. 

Each row of the table concerns a family or sub-family of rational
graphs, and each column corresponds to a structural restriction of
that family with respect to sets of initial vertices and degree. The
first case is that of rational (infinite) sets of initial vertices,
while the second case only considers the fixed rational initial set
$\{a\}^*$ over a single letter $a$. The two remaining cases concern
graphs with a unique initial vertex, with respectively arbitrary and
finite out-degree. A cell containing an equality symbol indicates that
the languages accepted by the considered family of graphs (row) from
the considered set of initial vertices (column) are the
context-sensitive languages. An inclusion symbol indicates that their
languages are strictly included in context-sensitive languages.  A
question mark denotes a conjecture.  When relevant, we give a
reference to the proposition, theorem or remark which states each
result.

\begin{table}[ht]
  \begin{center}
    \renewcommand{\arraystretch}{1.5}
    \begin{tabular}{l||c|c|c|c}
      \multirow{2}{*}{\bf Family of graphs} & \multicolumn{4}{c}{\bf
        Set of initial vertices} \\\cline{2-5} 
      & Rational set & \quad Set $\{a\}^*$ \quad 
      & Unique vertex & \parbox[c][7ex]{7em}{\centering 
        Unique vertex\\(finite degree)}
      \\
      \hline
      \hline
      Rational \cite{Morvan01a} & $=$ & $=$ & $=$ 
      & $=$ [\ref{th:cs-ratfd}]
      \\
      \hline
      Synchronized \cite{Rispal02} & $=$ & $=$ & 
      $=$ [\ref{lem:one-point}] & $\subset$
      % \footnotemark
      (?)  [\ref{thm:carac-sdb}]
      \\
      \hline
      Synchronous \cite{Rispal02}& $=$ & $=$ [\ref{th:cs-synch}]
      & $\subset$ [\ref{rem:one-point-synch}] & $\subset$
      \\
      \hline
      Sequential \hbox{synchronous} & $=$ [\ref{prop:cs-seq}]
      & $\subset$ (?) [\ref{prop:quitue}] & $\subset$ & $\subset$ 
      \\
      \hline
    \end{tabular}
  \end{center}
  \caption{Families of rational graphs and their languages.} 
  \label{tab:classif}
\end{table}

Finally, we investigate the case of deterministic languages. A
long-standing open problem in language theory is the equivalence
between deterministic and non-deterministic (or even unambiguous)
context-sensitive languages \cite{Kuroda64}. Thanks to our
constructions, we characterize two syntactical sub-families of
rational graphs respectively accepting the unambiguous and
deterministic context-sensitive languages.

{\bf Outline.} Our presentation is structured along the following
lines. The definitions of rational graphs and context-sensitive
languages are given in Section \ref{sect:defs}. The results concerning
languages accepted by rational and synchronous rational graphs are
given in Section \ref{sect:rat}. In Section \ref{sect:automata}, we
investigate rational graphs under structural constraints, and finally
Section \ref{sect:determinism} is devoted to deterministic
context-sensitive languages.

\section{Definitions}
\label{sect:defs}

\subsection{Notations}

Before all, we fix notations for words, languages and automata, as
well as directed graphs and the \intro{languages} they accept. For a
more thorough introduction to monoids and rationality, the interested
reader is referred to \cite{Berstel79,Sakarovitch03}.

\subsubsection{Languages and Automata}

We consider finite sets of symbols, or \intro{letters}, called
\intro{alphabets}. In the following, $\Sigma$ and $\Gamma$ always
denote finite alphabets. Tuples of letters are called \intro{words},
and sets of words \intro{languages}. The word $u$ corresponding to the
tuple $(u_1, \ldots, u_n)$ is written $u_1 \ldots u_n$. Its $i$-th
letter is denoted by $u(i) = u_i$. The set of all words over $\Sigma$
is written $\Sigma^*$. The number of letter occurrences of $u$ is its
length, written $|u|=n$. The unique word of length $0$ is written
$\eps$. The concatenation of two words $u = u_1 \ldots u_n$ and $v =
v_1 \ldots v_m$ is the word $uv = u_1 \ldots u_n v_1 \ldots v_m$. This
operation extends to sets of words: for all $A,B \subseteq \Sigma^*$,
$A B$ stands for the set $\{ uv \sep u \in A \; \textrm{and} \; v \in
B \}$. By a slight abuse of notation, we will usually denote by $u$
both the word $u$ and the singleton $\{u\}$.

A \intro{monoid} is composed of a set $M$ together with an associative
internal binary law on $M$ called \intro{product}, with a neutral
element in $M$. The product of two elements $x$ and $y$ of $M$ is
written $x \cdot y$. An automaton over $M$ is a tuple $\automate = (L,
Q, q_0, F, \delta)$ where $L \subseteq M$ is a finite set of
\intro{labels}, $Q$ a finite set of \intro{control states}, $q_0 \in
Q$ is the \intro{initial state}, $F \subseteq Q$ is the set of
\intro{final states} and $\delta \subseteq Q \times L \times Q$ is the
\intro{transition relation} of $A$. 

A \intro{run} of $\automate$ is a sequence of transitions
$(q_0,l_1,q_1) \ldots (q_{n-1},l_{n},q_n)$. It is associated to the
element $m = l_1 \cdot \ldots \cdot l_n \in M$.  If $q_n$ belongs to
$F$, the run is \intro{accepting} (or \intro{successful}), and $m$ is
accepted, or recognized, by $\automate$. The set of elements accepted
by $\automate$ is written $L(\automate)$. $\automate$ is
\intro{unambiguous} if there is only one accepting run for each
element in $L(\automate)$.

The star of a set $X \subseteq M$ is defined as $X^* := \bigcup_{k\geq
  0} X^k$ with $X^0 = \{\eps\}$ and $X^{k+1} = X \cdot X^k$.
Similarly, we write $X^+ := \bigcup_{k \geq 1} X^k$. The set of
\intro{rational} subsets of a monoid is the smallest set containing
all finite subsets and closed under union, product and star.  The set
of all words over $\Sigma$ together with the concatenation operation
forms the so-called \intro{free monoid} whose neutral element is the
empty word $\eps$. Finite automata over the free monoid $\Sigma^*$ are
known to accept the rational subsets of $\Sigma^*$, also called
rational languages.

\subsubsection{Graphs}

A labeled, directed and simple \intro{graph} is a set $G \subseteq V
\times \Gamma \times V$ where $\Gamma$ is a finite set of labels and
$V$ a countable set of \intro{vertices}. An element $(s,a,t)$ of $G$
is an \intro{edge} of \intro{source} $s$, \intro{target} $t$ and
\intro{label} $a$, and is written $s \era{a}{G} t$ or simply $s
\erb{a} t$ if $G$ is understood. The set of all sources and targets of
a graph form its \emph{support} $V_G$. A sequence of edges $s_1
\erb{a_1} t_1, \ldots, s_k \erb{a_k} t_k$ with $\forall i \in [2,k],\
s_i = t_{i-1}$ is called a \intro{path}. It is written $s_1 \erb{u}
t_k$, where $u = a_1 \ldots a_k$ is the corresponding \intro{path
  label}.  A graph is \intro{deterministic} if it contains no pair of
edges having the same source and label. The path language of a graph
$G$ between two sets of vertices $I$ and $F$ is the set
\[
L(G,I,F)\ :=\ \{\ w\ |\ s \era{w}{G} t,\ s \in I,\ t \in F \}.
\]
If two infinite automata recognize the same language, we say they are
\emph{trace-equivalent}. In this paper, we consider infinite automata:
infinite graphs together with sets of initial and final vertices. We
will no longer distinguish the notion of graph with initial and final
vertices from the notion of automaton. However, as we will see in
Section \ref{sect:rat}, with no restriction on the set of initial
vertices and on the structure of the graph this might not provide a
reasonable extension of finite automata.

\subsection{Word transducers}
\label{ssec:transd}

Automata can be used to accept more than languages. In particular,
when the edges of an automaton are labelled with pairs of letters
(with an appropriate product operation), its language is a set of
pairs of words, which can be seen as a binary relation on words. Such
automata are called finite automata with output, or
\intro{transducers}, and they recognize rational relations. We will
now recall their definition as well as some of their important
properties.  For a detailed presentation of transducers, see for
instance \cite{Berstel79,Prieur00,Sakarovitch03}.

Consider the monoid whose elements are the pairs of words $(u,v)$ in
$\Sigma^*$, and whose composition law is defined by $(u_1,v_1) \cdot
(u_2,v_2) = (u_1u_2, v_1v_2)$, generally called the product monoid and
written $\Sigma^* \times \Sigma^*$. A transducer $T$ over a finite
alphabet $\Sigma$ is a finite automaton over $\Sigma^* \times
\Sigma^*$ with labels in $(\Sigma \cup \{\eps\}) \times (\Sigma \cup
\{\eps\})$. Finite transducers accept the rational subsets of
$\Sigma^* \times \Sigma^*$. We do not distinguish a transducer from
the relation it accepts and write $(w,w') \in T$ if $(w,w')$ is
accepted by $T$. The \intro{domain} $\mathrm{Dom}(T)$ (resp.
\intro{range} $\mathrm{Ran}(T)$) of a transducer $T$ is the set $\{ w
\; | \; (w,w') \in T\}$ (resp. $\{ w' \; |\; (w,w') \in T \}$). We
also write $T(L)$ the set of all vertices $v$ such that $(u,v)\in T$
for some $u\in L$.  A transducer accepting a function is called
\intro{functional}.

In general, there is no bound on the size difference between input and
output in a transducer. Interesting subclasses are obtained by
enforcing some form of synchronization. For instance,
\emph{length-preserving} rational relations are recognized by
transducers with labels in $\Sigma \times \Sigma$, called
\emph{synchronous} transducers. Such relations only pair words of the
same size. 

A more relaxed form of synchronization was introduced by Elgot and
Mezei \cite{Elgot65}: a transducer over $\Sigma$ with initial state
$q_0$ is left-synchronized if for every path
\[
q_0 \erb{x_0/y_0} q_1 \ldots q_{n-1} \erb{x_n/y_n} q_n,
\]
there exists $k \in [0,n]$ such that for all $i \in [0,k]$, $x_i$ and
$y_i$ belong to $\Sigma$ and either $x_{j}=\eps$ for all $j > k$ or
$y_{j} = \eps$ for all $j > k$. In other terms, a left-synchronized
relation is a finite union of relations of the form $S \cdot F$ where
$S$ is a synchronous relation and $F$ is either equal to $\{ \eps \}
\times R$ or $R \times \{ \eps \}$ where $R$ is a rational language.
\intro{Right-synchronized} transducers are defined similarly. In the
following, unless otherwise stated, we will refer to left-synchronized
transducers simply as synchronized transducers.

The standard notion of determinism for automata does not have much
meaning in the case of transducers because it does not rely only on
the input but on both the input and the output. A more refined notion
is that of sequentiality: a transducer $T$ with states $Q$ is
\intro{sequential} if for all $q,q'$ and $q''$ in $Q$, if $q \erb{x/y}
q'$ and $q \erb{x'/y'} q''$ then either $x=x'$, $y=y'$ and $q'=q''$,
or $x \neq \varepsilon$, $x'\neq \varepsilon$ and $x \neq x'$.

\begin{rem}
\label{rem:sync-unam}
The standard determinization procedure applied to a synchronous
transducer yields an equivalent unambiguous synchronous transducer
(i.e for every pair of words $(u,v)$ accepted by the transducer there
is exactly one accepting run of the transducer labelled by $u/v$).
This remains true for synchronized transducers.
\end{rem}

It is well-known that there is a close relationship between rational
languages and rational transductions. In particular, rational
relations have rational domains and ranges, and are closed under
restriction to a rational domain or range.  Moreover, the restriction
of a sequential (resp. synchronous) transducer to a rational domain is
still sequential (resp.  synchronous) (see for instance
\cite{Berstel79}).

\subsection{Rational graphs}

The Chomsky-like hierarchy of graphs presented in \cite{Caucal02} uses
words to represent vertices. Each of these graphs is thus a finite
union of binary relations on words, each relation corresponding to a
given edge label. In particular, the family of rational graphs owes
its name to the fact that their sets of edges are given by rational
relations on words, i.e. relations recognized by word transducers.

\begin{defi}[\cite{Morvan00}]
  A rational graph labelled by $\Sigma$ with vertices in $\Gamma^*$ is
  given by a tuple of transducers $(T_a)_{a \in \Sigma}$ over
  $\Gamma$. For all $a \in \Sigma$, $G$ has an edge labelled by $a$
  between vertices $u$ and $v \in \Gamma^*$ if and only if $(u,v) \in
  T_a$. For $w \in \Sigma^+$ and $a \in \Sigma$, we write $T_{wa} =
  T_w \circ T_a$, and $u \erb{w} v$ if and only if $(u,v) \in T_w$.
\end{defi}

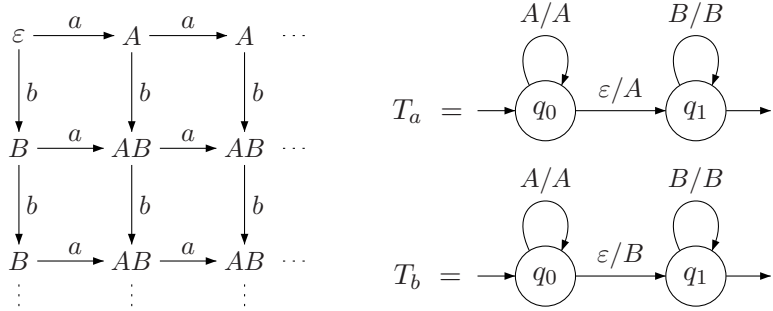
\begin{figure}
\begin{center}
\begin{picture}(100,40)(0,-36)
  \gasset{loopdiam=6}

  \node[Nframe=n](ta)(55,-10){$T_a\ =\ $}
  \node[Nframe=n](tb)(55,-32){$T_b\ =\ $}
  \node[Nmarks=i,ilength=5](q0)(70,-10){$q_0$}
  \node[Nmarks=i,ilength=5](q00)(70,-32){$q_0$}
  \node(q1)(90,-10){$q_1$}
  \node(q2)(90,-32){$q_1$}
  \node[Nframe=n,Nw=0](q11)(100,-10){}
  \node[Nframe=n,Nw=0](q22)(100,-32){}

  \drawloop(q0){\small $A / A$}
  \drawloop(q00){\small $A / A$}
  \drawedge(q0,q1){\small $\eps / A$}
  \drawedge(q00,q2){\small $\eps / B$}
  \drawedge(q1,q11){}
  \drawedge(q2,q22){}
  \drawloop(q1){\small $B / B$}
  \drawloop(q2){\small $B / B$}

  \gasset{Nframe=n,Nadjust=wh}

  \node(A1)( 0,0){$\eps$}
  \node(A2)(15,0){\small $A$}
  \node(A3)(30,0){\small $A²$}

  \node(B1)( 0,-15){\small $B$}
  \node(B2)(15,-15){\small $AB$}
  \node(B3)(30,-15){\small $A²B$}

  \node(C1)( 0,-30){\small $B²$}
  \node(C2)(15,-30){\small $AB²$}
  \node(C3)(30,-30){\small $A²B²$}

  \drawedge(A1,A2){\small $a$}
  \drawedge(A2,A3){\small $a$}
  \drawedge(B1,B2){\small $a$}
  \drawedge(B2,B3){\small $a$}
  \drawedge(C1,C2){\small $a$}
  \drawedge(C2,C3){\small $a$}

  \drawedge(A1,B1){\small $b$}
  \drawedge(A2,B2){\small $b$}
  \drawedge(A3,B3){\small $b$}
  \drawedge(B1,C1){\small $b$}
  \drawedge(B2,C2){\small $b$}
  \drawedge(B3,C3){\small $b$}

  \gasset{AHnb=0,dash={0.3 1}0}
  \drawline(35,0)(38,0)
  \drawline(35,-15)(38,-15)
  \drawline(35,-30)(38,-30)
  \drawline(0,-33)(0,-36)
  \drawline(15,-33)(15,-36)
  \drawline(30,-33)(30,-36)
\end{picture}
\end{center}

\caption{The grid and its associated transducers.} 
\label{fig:grid}
\end{figure}

Note that $T_w \circ T_a$ stands for the set of all pairs $(u,v)$ such
that $(u,x)\in T_w$ and $(x,v)\in T_a$ for some $x$. Figure
\ref{fig:grid} shows an example of rational graph, the infinite
\emph{grid}, with the rational transducers which define its edges.  By
the properties of rational relations, the support of a rational graph
is a rational subset of $\Gamma^*$. The rational graphs with
synchronized transducers were already defined by Blumensath and Grädel
in \cite{Blumensath00} under the name \intro{automatic} graphs and by
Rispal in \cite{Rispal02} under the name \intro{synchronized} rational
graphs.

It follows from the definitions (see Section \ref{ssec:transd}) that
sequential synchronous, synchronous, synchronized and rational graphs
form an increasing hierarchy. This hierarchy is strict (up to
isomorphism): first, sequential synchronous graphs are deterministic
graphs whereas synchronous graphs can be non-deterministic. Second,
synchronous graphs have a finite degree whereas synchronized graphs
can have an infinite degree. Finally, to separate synchronized
graphs from rational graphs, we can use the following properties on
the growth rate of the out-degree in the case of graphs of finite
out-degree. 

\begin{prop}{\cite{Morvan01b}}
  For any rational graph $G$ of finite out-degree and any vertex $x$,
  there exists $c \in \mathbb{N}$, such that the out-degree of
  vertices at distance $n$ of $x$ is at most $c^{c^n}$.
\end{prop}

This upper bound can be reached: consider the unlabeled rational graph
$G_0=\{T\}$ where $T$ is the transducer over $\Gamma=\{A,B\}$ with one
state $q_0$ which is both initial and final and a transition $q_0
\erb{X/YZ} q_0$ for all $X, Y$ and $Z \in \Gamma$.  It has an
out-degree of $2^{2^{n+1}}$ at distance $n$ of $A$.  In the case of
synchronized graphs of finite out-degree, the bound on the out-degree
is simply exponential.

\begin{prop}{\cite{Rispal02}}
  For any synchronized graph $G$ of finite out-degree and vertex $x$,
  there exists $c \in \mathbb{N}$ such that the out-degree of
  vertices at distance $n>0$ of $x$ is at most $c^n$.
\end{prop}

It follows from the above proposition that $G_0$ is rational but not
synchronized. Hence, the synchronized graphs form a strict sub-family
of rational graphs.

\subsection{Context sensitive languages}

In this work, we are concerned with the family of
\intro{context-sensitive languages}\footnote{In order to simplify our
  presentation, we only consider context-sensitive languages that do
  not contain the empty word $\eps$ (this is a standard
  restriction).}. Several finite formalisms are known to accept this
family of languages, the most common being linearly bounded machines
(LBM), which are Turing machines working in linear space. Less
well-known acceptors for these languages are bounded tiling systems,
which are not traditionally studied as language recognizers. However,
one can show that these formalisms are equivalent, and that
syntactical translations exist between them.  Since they are at the
heart of our proof techniques, we now give a detailed definition of
tiling systems. For more information about linearly bounded machines
the reader is referred to \cite{Hopcroft79}.

Tiling systems were originally defined to recognize or specify
\intro{picture languages}, i.e. two-dimensional words on finite
alphabets \cite{Giammarresi96}. They can be seen as a normalized form
of dominos systems \cite{LS97}. Such sets of pictures are called
\intro{local} picture languages. However, by only looking at the words
contained in the first row of each picture of a local picture
language, one obtains a context-sensitive language \cite{Latteux97}.

A $(n,m)$-picture $p$ over an alphabet $\Gamma$ is a two dimensional
array of letters in $\Gamma$ with $n$ rows and $m$ columns. We denote
by $p(i,j)$ the letter occurring in the $i$th row and $j$th column
starting from the top-left corner, by $\Gamma^{n,m}$ the set of
$(n,m)$-pictures and by $\Gamma^{**}$ the set of all
pictures\footnote{We do not consider the empty picture.}. Given a
$(n,m)$-picture $p$ over $\Gamma$ and a letter $\sd \not \in \Gamma$,
we denote by ${p}_{\td}$ the $(n+2,m+2)$-picture over $\Gamma \cup
\{\#\}$ defined by:
\begin{itemize}
\item ${p}_{\td}(i,1)={p}_{\td}(i,m+2)=\sd$ for $i\in [1,n+2]$,
\item ${p}_{\td}(1,j)={p}_{\td}(n+2,j)=\sd$ for $j\in [1,m+2]$,
\item ${p}_{\td}(i+1,j+1)=p(i,j)$ for $i \in [1,n]$ and $j \in [1,m]$. 
\end{itemize}
For any $n,m \geq 2$ and any $(n,m)$-picture $p$, $T(p)$ is the set of
$(2,2)$-pictures appearing in $p$. A $(2,2)$-picture is also called a
\emph{tile}. A picture language $K \subseteq \Gamma^{**}$ is
\emph{local} if there exists a symbol $\sd \not\in \Gamma$ and a
finite set of tiles $\Delta$ such that $K = \{ p \in \Gamma^{**} \mid
T({p}_{\td}) \subseteq \Delta \}$. To any set of pictures over
$\Gamma$, we can associate a language of words by looking at the
frontiers of the pictures. The frontier of a $(n,m)$-picture $p$ is
the word $\fr{p}=p(1,1) \ldots p(1,m)$ corresponding to the first row
of the picture.

\begin{defi}
  A tiling system $S$ is a tuple $(\Gamma, \Sigma, \sd, \Delta)$ where
  $\Gamma$ is a finite alphabet, $\Sigma \subset \Gamma$ is the input
  alphabet, $\sd \not \in \Gamma$ is a frame symbol and $\Delta$ is a
  finite set of tiles over $\Gamma \cup \{\sd\}$. It recognizes the
  local picture language $P(S) = \{ p \in \Gamma^{**} \mid
  T({p}_{\td}) \subseteq \Delta \}$ and the word language $L(S) = \fr{
    P(S)} \cap \Sigma^*$.
\end{defi}

\begin{figure}
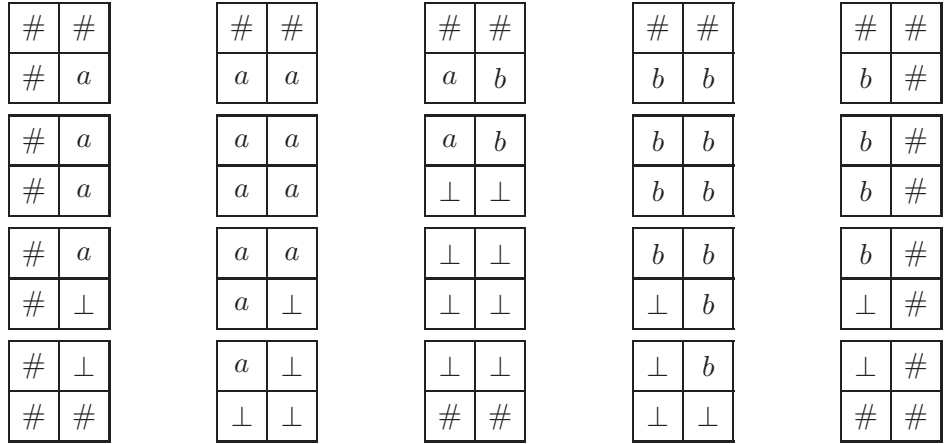

  \begin{center}
    \begin{align*}
      \stile{\#}{\#}{\#}{a} && \stile{\#}{\#}{a}{a} &&
      \stile{\#}{\#}{a}{b} && \stile{\#}{\#}{b}{b} &&
      \stile{\#}{\#}{b}{\#}
      \\
      \stile{\#}{a}{\#}{a} && \stile{a}{a}{a}{a} &&
      \stile{a}{b}{\bot}{\bot} && \stile{b}{b}{b}{b} &&
      \stile{b}{\#}{b}{\#}
      \\
      \stile{\#}{a}{\#}{\bot} && \stile{a}{a}{a}{\bot} &&
      \stile{\bot}{\bot}{\bot}{\bot} && \stile{b}{b}{\bot}{b} &&
      \stile{b}{\#}{\bot}{\#}
      \\
      \stile{\#}{\bot}{\#}{\#} && \stile{a}{\bot}{\bot}{\bot} &&
      \stile{\bot}{\bot}{\#}{\#} && \stile{\bot}{b}{\bot}{\bot} &&
      \stile{\bot}{\#}{\#}{\#}
    \end{align*}
  \end{center}
  \caption{A tiling system accepting $a^nb^n$ (Cf. Example
    \ref{ex:ts}).} 
  \label{fig:ts}
\end{figure}

A tiling system $S$ recognizes a language $L \subseteq \Sigma^+$ in
height $f(n)$ for some mapping $f: \mathbb{N} \mapsto \mathbb{N}$ if
for all $w \in L(S)$ there exists a $(n,m)$-picture $p$ in $P(S)$ such
that $w=\fr{p}$ and $n \leq f(m)$. We can now precisely state the
following well-known equivalence result.

\begin{thm}
  \label{thm:equiv-recogn}
  The following simulations link linearly bounded machines and tiling
  systems:
  \begin{enumerate}
  \item A linearly bounded Turing machine ${T}$ working in $f(n)$
    reversals can be simulated by a tiling system of height $f(n)+2$.
  \item A tiling system of height $f(n)$ can be simulated by a
    linearly bounded Turing machine working in $f(n)$ reversals.
  \end{enumerate}
\end{thm}

\begin{exa}
  \label{ex:ts}
  Figure \ref{fig:ts} shows the set of tiles $\Delta$ of a tiling
  system $S$ over $\Gamma = \{a,b,\bot\}$, $\Sigma = \{a,b\}$ and the
  border symbol $\sd$. The language $L(S)$ is exactly the set $\{ a^n
  b^n \; | \; n \geq 1 \}$.
\end{exa}

A context-sensitive language is called \intro{deterministic} if it can
be accepted by a deterministic LBM or tiling system, where a tiling
system is deterministic if one can infer from each row in a picture a
single possible next row.

\section{The languages of rational graphs}
\label{sect:rat}

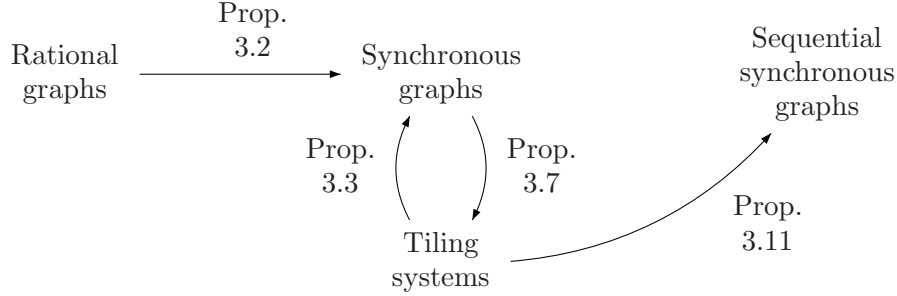
\begin{figure}
  \begin{center}
    \begin{picture}(100,32)(0,-27)
      \gasset{Nframe=n,NLangle=0.0,Nadjust=wh,Nmr=0.0}
      \node[Nmr=1.95](n0)(0,0){\begin{tabular}{c}Rational\\graphs\end{tabular}}

      \node(n1)(50,0){\begin{tabular}{c}Synchronous\\graphs\end{tabular}}
      \node(n2)(100,0){\begin{tabular}{c}Sequential\\synchronous\\graphs\end{tabular}}
      \node(n3)(50,-25){\begin{tabular}{c}Tiling\\systems\end{tabular}}

      \drawedge(n0,n1){\begin{tabular}{c}Prop.\\\ref{prop:rat2synch}\end{tabular}}
      \drawedge[ELpos=70,ELside=r,curvedepth=-8](n3,n2){\begin{tabular}{c}Prop.\\\ref{prop:cs-seq}\end{tabular}}

      \gasset{curvedepth=6}
      \drawedge(n1,n3){\begin{tabular}{c}Prop.\\\ref{prop:synch2cs}\end{tabular}}
      \drawedge(n3,n1){\begin{tabular}{c}Prop.\\\ref{prop:cs2synch}\end{tabular}}
    \end{picture}
  \end{center}
  \caption{Each edge represents an effective transformation preserving
    languages.}
  \label{fig:scheme}
\end{figure}

In this section, we consider the languages accepted by rational graphs
and their sub-families from and to a rational set of vertices. We give
a simplified presentation of the result by Morvan and Stirling
\cite{Morvan01a} stating that the family of rational graphs accepts
the context-sensitive languages. This is done in several steps. First,
Proposition~\ref{prop:rat2synch} states that the rational graphs are
trace-equivalent to the \emph{synchronous} rational graphs. Then,
Proposition \ref{prop:cs2synch} and Proposition \ref{prop:synch2cs}
establish a very tight relationship between synchronous graphs and
tiling systems.  It follows that the languages of synchronous rational
graphs are also the context-sensitive languages (Theorem
\ref{th:cs-synch}). The original result is given as Corollary
\ref{cor:cs-rat}. Finally, Proposition \ref{prop:cs-seq} establishes
that even the smallest sub-family we consider, the family of
sequential synchronous rational graphs, accepts all context-sensitive
languages. The various transformations presented in this section are
summarized in Figure~\ref{fig:scheme}.

\subsection{From rational graphs to synchronous graphs}

We present an effective construction that transforms a rational graph
$G$ with two rational sets $I$ and $F$ of initial and final vertices
into a synchronous graph $G'$ trace-equivalent between two rational
sets $I'$ and $F'$. The construction is based on replacing the symbol
$\eps$ in the transitions of the transducers defining $G$ by a fresh
symbol $\sd$.

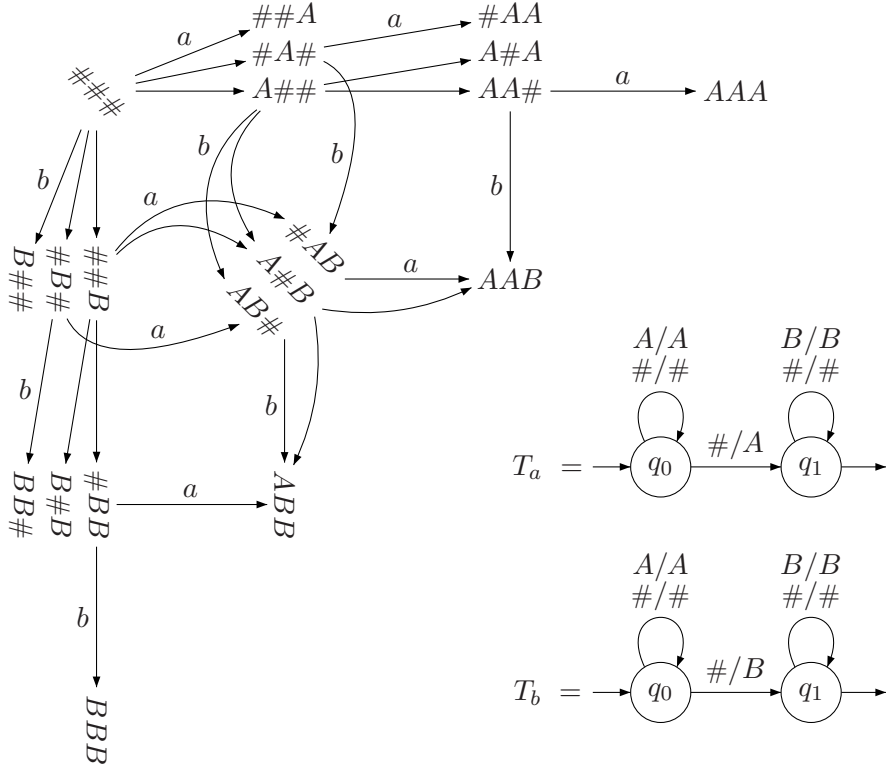
\begin{figure}
\begin{center}
\begin{picture}(120,96)(-8,-92)
  \gasset{loopdiam=6}
  \node[Nframe=n](ta)(65,-55){$T_a\ =$}
  \node[Nframe=n](tb)(65,-85){$T_b\ =$}
  \node[Nmarks=i,ilength=5](q0)(80,-55){$q_0$}
  \node[Nmarks=i,ilength=5](q00)(80,-85){$q_0$}
  \node(q1)(100,-55){$q_1$}
  \node(q2)(100,-85){$q_1$}
  \node[Nframe=n,Nw=0](q11)(110,-55){}
  \node[Nframe=n,Nw=0](q22)(110,-85){}

  \renewcommand{\arraystretch}{0.5}

  \drawloop(q0){$\begin{array}{c}A / A \\ \sd / \sd\end{array}$}
  \drawloop(q00){$\begin{array}{c}A / A \\ \sd / \sd\end{array}$}
  \drawedge(q0,q1){$\sd / A$}
  \drawedge(q00,q2){$\sd / B$}
  \drawedge(q1,q11){}
  \drawedge(q2,q22){}
  \drawloop(q1){$\begin{array}{c}B / B \\ \sd / \sd\end{array}$}
  \drawloop(q2){$\begin{array}{c}B / B \\ \sd / \sd\end{array}$}

  \renewcommand{\arraystretch}{1.0}

  \gasset{Nframe=n,Nadjust=wh,Nw=10,Nh=6}

  \node(ddd)(5,-5){\begin{turn}{-45}$\sd \sd \sd$\end{turn}}

  \node(add)(30,-5){$A   \sd \sd$}
  \node(dad)(30, 0){$\sd A   \sd$}
  \node(dda)(30, 5){$\sd \sd A$}

  \node(ddb)( 5,-30){\begin{turn}{-90}$\sd \sd B$\end{turn}}
  \node(dbd)( 0,-30){\begin{turn}{-90}$\sd B   \sd$\end{turn}}
  \node(bdd)(-5,-30){\begin{turn}{-90}$B   \sd \sd$\end{turn}}

  \node[Nadjust=n,Nw=0,Nh=0](abd)(26,-34){\begin{turn}{-45}$A B \sd$\end{turn}}
  \node[Nadjust=n,Nw=0,Nh=0](adb)(30,-30){\begin{turn}{-45}$A \sd B$\end{turn}}
  \node[Nadjust=n,Nw=0,Nh=0](dab)(34,-26){\begin{turn}{-45}$\sd A B$\end{turn}}

  \node(aad)(60,-5){$A A \sd$}
  \node(ada)(60,0){$A \sd A$}
  \node(daa)(60,5){$\sd A A$}

  \node(dbb)( 5,-60){\begin{turn}{-90}$\sd B B$\end{turn}}
  \node(bdb)( 0,-60){\begin{turn}{-90}$B \sd B$\end{turn}}
  \node(bbd)(-5,-60){\begin{turn}{-90}$B B \sd$\end{turn}}

  \node(aaa)(90,-5){$A A A$}
  \node(aab)(60,-30){$A A B$}
  \node(abb)(30,-60){\begin{turn}{-90}$A B B$\end{turn}}
  \node(bbb)( 5,-90){\begin{turn}{-90}$B B B$\end{turn}}

  \drawedge(ddd,add){}
  \drawedge(ddd,dad){}
  \drawedge(ddd,dda){$a$}

  \drawedge[ELside=r](ddd,bdd){$b$}
  \drawedge(ddd,dbd){}
  \drawedge(ddd,ddb){}

  \drawedge(add,aad){}
  \drawedge(add,ada){}
  \drawedge(dad,daa){$a$}

  \drawedge[ELside=r](dbd,bbd){$b$}
  \drawedge(ddb,dbb){}
  \drawedge(ddb,bdb){}

  \drawedge(aad,aaa){$a$}

  \drawedge[ELside=r](dbb,bbb){$b$}

  \drawedge[exo=-4,eyo=5,curvedepth=-5](add,adb){}
  \drawedge[exo=-4,eyo=4,curvedepth=-6,ELside=r,ELpos=40](add,abd){$b$}
  \drawqbedge[exo=2,eyo=2,ELside=r,ELpos=75](dad,45,0,dab){$b$}

  \drawqbedge[exo=-2,eyo=-2,ELpos=75](dbd,0,-45,abd){$a$}
  \drawedge[exo=-5,eyo=4,curvedepth=5](ddb,adb){}
  \drawedge[exo=-4,eyo=4,curvedepth=6,ELpos=40](ddb,dab){$a$}

  \drawedge[ELside=r](aad,aab){$b$}
  \drawedge[exo=-3,sxo=5,syo=-4,curvedepth=-2](adb,aab){}
  \drawedge[sxo=4,syo=-4,ELpos=40](dab,aab){$a$}
  \drawedge[eyo=3,sxo=4,syo=-5,curvedepth=2](adb,abb){}
  \drawedge[sxo=4,syo=-4,ELside=r,ELpos=40](abd,abb){$b$}
  \drawedge(dbb,abb){$a$}

\end{picture}
\end{center}

\caption{Synchronous graph trace-equivalent to the grid (1 connected
  component).} 
\label{fig:modgrid}
\end{figure}

Let $(T_a)_{a \in \Sigma}$ be the set of transducers over $\Gamma$
characterizing $G$ and let $\sd$ be a symbol not in $\Gamma$.  For all
$a$, we define $\bar{a}$ to be equal to $a$ if $a \in \Gamma$, and to
$\eps$ if $a = \sd$. We extend this to a projection from $(\Gamma \cup
\sd)^*$ to $\Gamma^*$ in the standard way. We define $G'$ as the
rational graph defined by the set of transducers $(T'_a)_{a \in
  \Sigma}$ where $T'_a$ has the same set of control states $Q_a$ as $T_a$ and a
set of transitions given by
\[
\big\{ \, p \erb{a/b} q \sep p \erb{\bar{a}/\bar{b}} q \in T_a \,
\big\}\ \cup\ \big\{ \, p \erb{\td/\td} p \sep p \in Q_a \, \big\}.
\]
By definition of each $T'_a$, $G'$ is a synchronous rational graph.
Let $I'$ and $F'$ be the two rational sets such that $I'=\{ u \; | \;
\bar{u} \in I\}$ and $F'=\{ v \; | \; \bar{v} \in F \}$ (the automaton
accepting $I'$ (resp. $F'$) is obtained from the automaton accepting
$I$ (resp.  $F$) by adding a loop labeled by $\sd$ on each control
state). We claim that $G'$ accepts between $I'$ and $F'$ the same
language as $G$ between $I$ and $F$. For example, Figure
\ref{fig:modgrid} illustrates the previous construction applied to the
graph of Figure \ref{fig:grid}.  Only one connected component of the
obtained graph is shown.

Before we prove the correctness of this construction, we need to
establish a couple of technical lemmas. Let $\mathcal{B}$ be the set
of all mappings from $\mathbb{N}$ to $\mathbb{N}$.  To any mapping
$\delta \in \mathcal{B}$, we associate a mapping from $(\Gamma \cup
\{\sd\})^*$ to $(\Gamma \cup \{\sd\})^*$ defined as follows: for all
$w = \sd^{i_0} a_1 \sd^{i_1} \ldots a_n \sd^{i_n}$ with $a_1, \ldots,
a_n \in \Gamma$, let $\delta w = \sd^{i_0 + \delta(0)} a_1 \sd^{i_1 +
  \delta(1)} \ldots a_n \sd^{i_n + \delta(n)}$. Before proceeding, we
state two properties of these mappings with respect to the sets of
transducers $(T_a)$ and $(T'_a)$.

\begin{lem}
  \label{lem:buff}
  We have the following properties:
  \begin{align}
    \label{eq:buff1}
    \forall u,v \in \Gamma^*,\ \, (u,v) \in T_a\ \iff\ \exists
    \delta_u, \delta_v \in \mathcal{B},& \ (\delta_u u, \delta_v v)
    \in T'_a,
    \\
    \label{eq:buff2}
    \forall (u,v) \in T'_a,\ \forall \delta \in \mathcal{B},\ \exists
    \delta' \in \mathcal{B},&\ (\delta u, \delta' v) \in T'_a,
    \\
    \label{eq:buff3}
    \text{and dually \quad } \forall (u,v) \in T'_a,\ \forall \delta
    \in \mathcal{B},\ \exists \delta' \in \mathcal{B},&\ (\delta' u,
    \delta v) \in T'_a.
  \end{align}
\end{lem}

We can now prove the correctness of the construction: a word $w$ is
accepted by $G$ between $I$ and $F$ if and only if it is accepted by
$G'$ between $I'$ and $F'$. 

\begin{prop}
  \label{prop:rat2synch}
  For every rational graph $G$ and rational sets of vertices $I$ and
  $F$, there is a synchronous graph $G'$ and two rational sets $I'$
  and $F'$ such that $L(G,I,F) = L(G',I',F')$.
\end{prop}

\begin{proof}
We show by induction on $n$ that for all
$u_0,\ldots,u_n \in \Gamma^*$, if there is a path
\[
u_0 \era{w(1)}{G} u_1 \ldots u_{n-1} \era{w(n)}{G} u_n,
\]
then there exist words $u'_0,\ldots,u'_n \in (\Gamma \cup \{\sd\})^*$
such that for all $i$, $\bar{u}'_i = u_i$, and
\[
u'_0 \era{w(1)}{G'} u'_1 \ldots u'_{n-1} \era{w(n)}{G'} u'_n. 
\]
The case where $n = 0$ is trivial. Suppose the property is true for
all paths of length at most $n$, and consider a path
\[
u_0 \era{w(1)}{G} \ldots \era{w(n)}{G} u_n \era{w(n+1)}{G} u_{n+1}. 
\]
By induction hypothesis, one can find mappings $\delta_0, \ldots
\delta_n$ such that
\[
\delta_0 u_0 \era{w(1)}{G'} \ldots \era{w(n)}{G'} \delta_n u_n. 
\]
We now use the properties of mappings stated in Lemma \ref{lem:buff}.
By (\ref{eq:buff1}), there exist $\delta'_n$ and $\delta'_{n+1}$ such
that $\delta'_n u_n \erb{w(n+1)} \delta'_{n+1} u_{n+1} \in G'$.  Let
$\gamma_n$ and $\gamma'_n$ be two elements of $\mathcal{B}$ such that
$\delta'_n \circ \gamma'_n = \delta_n \circ \gamma_n$. By Lemma
(\ref{eq:buff2}) and (\ref{eq:buff3}), we can find mappings
$\gamma'_{n+1}$ and $\gamma_0$ to $\gamma_{n-1}$ such that:
\[
\gamma_0 \delta_0 u_0 \era{w(1)}{G'} \ldots \era{w(n)}{G'} \gamma_n
\delta_n u_n = \gamma'_n \delta'_n u_n \era{w(n+1)}{G'} \gamma'_{n+1}
\delta'_{n+1} u_{n+1}
\]
which concludes the proof by induction. If we suppose that $u_0 \in I$
and $u_n \in F$, then necessarily $u'_0 \in I'$ and $u'_n \in F'$. It
follows that for every path in $G$ between $I$ and $F$, there is a path
in $G'$ between $I'$ and $F'$ with the same path label. 
  
Conversely, by (\ref{eq:buff1}), for any such path in $G'$, erasing
the occurrences of $\sd$ from its vertices yields a valid path in $G$
between $I$ and $F$.  Hence $L(G,I,F) = L(G',I',F')$.
\end{proof}

\subsection{Equivalence between synchronized graphs and tiling systems}

The following propositions establish the tight relationship between
tiling systems and synchronous rational graphs.  Proposition
\ref{prop:cs2synch} presents an effective transformation of a tiling
system into a synchronous rational graph. 

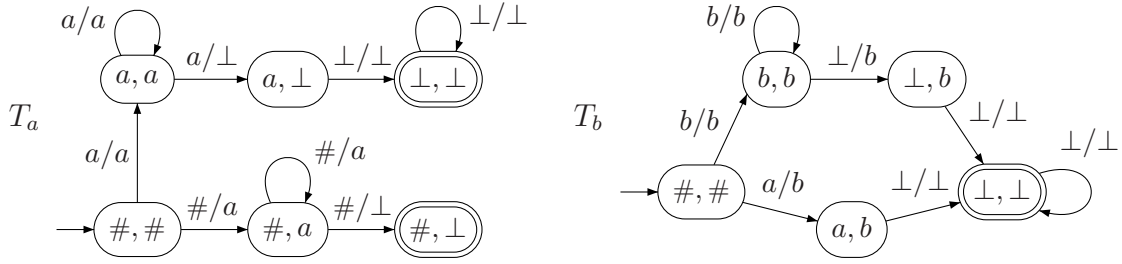
\begin{figure}
\begin{center}
\begin{picture}(120,36)(-60,0)
  \gasset{Nadjust=wh,Nadjustdist=2,loopdiam=6}

  \put(-65,3){
    \node[Nframe=n](ta)(-5,15){\large{$T_a$}}

    \node[Nmarks=i](dda)(10,0){$\sd,\sd$}
    \node(daa)(30,0){$\sd,a$}
    \node(aaa)(10,20){$a,a$}
    \node[Nmarks=r](dBa)(50,0){$\sd,\bot$}
    \node(aBa)(30,20){$a,\bot$}
    \node[Nmarks=r](BBa)(50,20){$\bot,\bot$}

    \drawedge(dda,daa){$\sd/a$}
    \drawedge(daa,dBa){$\sd/\bot$}

    \drawedge(dda,aaa){$a/a$}
    \drawedge(aaa,aBa){$a/\bot$}
    \drawedge(aBa,BBa){$\bot/\bot$}

    \drawloop[ELpos=70](daa){$\sd/a$}
    \drawloop[ELpos=25](aaa){$a/a$}
    \drawloop[ELpos=75](BBa){$\bot/\bot$}
  }

  \put(10,3){
    \node[Nframe=n](tb)(-5,15){\large{$T_b$}}

    \node[Nmarks=i](ddb)(10,5){$\sd,\sd$}
    \node(abb)(30,0){$a,b$}
    \node(bbb)(20,20){$b,b$}
    \node(Bbb)(40,20){$\bot,b$}
    \node[Nmarks=r](BBb)(50,5){$\bot,\bot$}

    \drawedge(ddb,abb){$a/b$}
    \drawedge(abb,BBb){$\bot/\bot$}
    \drawedge[exo=-3](ddb,bbb){$b/b$}
    \drawedge(bbb,Bbb){$\bot/b$}
    \drawedge(Bbb,BBb){$\bot/\bot$}

    \drawloop[loopangle=0,ELpos=30](BBb){$\bot/\bot$}
    \drawloop[ELpos=25](bbb){$b/b$}
  }

\end{picture}
\end{center}
\caption{Transducers of a synchronous graph accepting $\{ a^nb^n \mid
  n \geq 1 \}$.}
\label{fig:ts2rat-trans}
\end{figure}

\begin{prop}
  \label{prop:cs2synch}
  Given a tiling system $S=(\Gamma,\Sigma,\#,\Delta)$, there exists a
  synchronous rational graph $G$ and two rational sets $I$ and $F$
  such that $L(G,I,F)={L}(S)$. 
\end{prop}

\begin{proof}
  Consider the finite automaton $A$ on $\Gamma$ with a set of states
  $Q=\Gamma \cup \{\sd\}$, initial state $\sd$, a set of final states
  $F$ and a set of transitions $\delta$ given by:

  \[
  F\ :\ a \quad \text{ such that } \quad \stile{a}{\sd}{\sd}{\sd}\ \in
  \Delta
  \]

  \[
  \delta\ :\ \sd \era{a}{A} a, \ a \era{b}{A} b \text{\quad for all
    \quad} \stile{\sd}{\sd}{a}{\sd}, \quad \stile{a}{\sd}{b}{\sd} \in
  \Delta \text{ (respectively).}
  \]
  \smallskip

  \noindent
  Call $M$ the language recognized by $A$, $M$ represents the set of
  possible last columns of pictures of ${P}(S)$. Note that this does
  not imply that each word of $M$ actually \emph{is} the last column
  of a picture in ${P}(S)$, only that it is compatible with the right
  border tiles of $\Delta$.

  Let us build a synchronous rational graph $G$ and two rational sets
  $I$ and $F$ such that $L(G,I,F) = L(S)$. The transitions of the set
  of transducers $(T_e)_{e \in \Sigma}$ of $G$ are:
  \begin{align*}
    (\sd,\sd) & \era{c/d}{T_d} (c,d) & \text{\quad for all \quad}
    \stile{\sd}{\sd}{c}{d} & \in \Delta,\ d \neq \sd
    \\
    (a,b) & \era{c/d}{T_e} (c,d) & \text{\quad for all \quad}
    \stile{a}{b}{c}{d} & \in \Delta,\ b,d \neq \sd,\ e \in \Sigma
  \end{align*}
  where $(\sd,\sd)$ is the unique initial state of each transducer and
  the set of final states $F$ of each transducer is given by:
  \[
  \begin{array}{lcl}
    F & : &  (a,b) \in (\Gamma \cup \{ \sd \})  \times \Gamma \;\;
    \text{such that}\;\; \stile{a}{b}{\sd}{\sd}  \in \Delta.
  \end{array}
  \]
  A pair of words $(s,t)$ is accepted by the transducer $T_e$ if and
  only if $e$ is the first letter of $t$, and either $s$ and $t$ are
  two adjacent columns of a picture in ${P}(S)$ or $s \in \sd^*$ and
  $t$ is the first column of a picture in ${P}(S)$. As a consequence,
  $L(S) = L(G, \sd^*, M)$.
\end{proof}

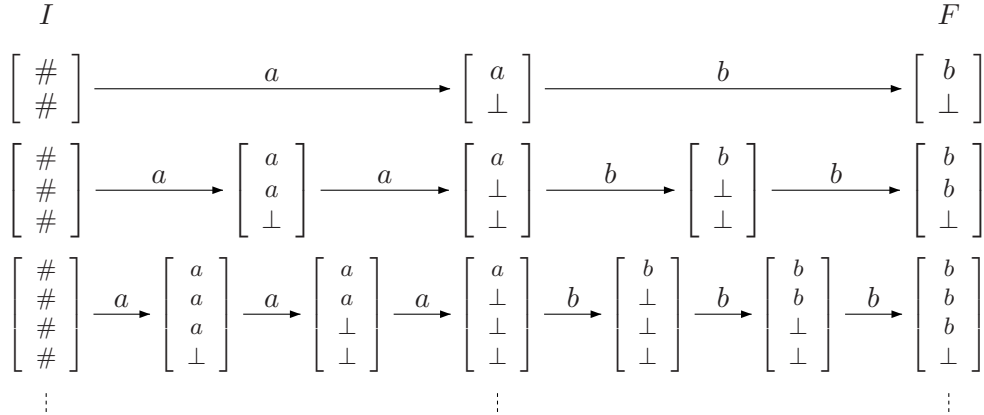
\begin{figure}
\begin{center}
  \begin{picture}(150,50)(0,-40)
\gasset{Nadjust=wh,Nadjustdist=1, loopdiam=7,Nframe=n,NLangle=0.0,Nmr=0.0}

\node(n00)(20.0,10.0){$I$}
\node(n00)(140.0,10.0){$F$}

\node(n20)(20.0,0){$\left[ \begin{array}{c} \# \\ \#  \end{array}\right]$}
\node(n21)(80.0,0){$\left[ \begin{array}{c} a \\ \bot \end{array}\right]$}
\node(n22)(140.0,0){$\left[ \begin{array}{c} b \\ \bot \end{array}\right]$}

\drawedge(n20,n21){$a$}
\drawedge(n21,n22){$b$}

{\small
\node(n40)(20.0,-13.5){$\left[ \begin{array}{c} \# \\ \#  \\ \# \end{array}\right]$}
\node(n41)(50.0,-13.5){$\left[ \begin{array}{c} a \\ a  \\ \bot \end{array}\right]$}
\node(n42)(80.0,-13.5){$\left[ \begin{array}{c} a \\ \bot \\ \bot \end{array}\right]$}
\node(n43)(110.0,-13.5){$\left[ \begin{array}{c} b \\ \bot \\ \bot \end{array}\right]$}
\node(n44)(140.0,-13.5){$\left[ \begin{array}{c} b \\ b \\ \bot \end{array}\right]$}}

\drawedge(n40,n41){$a$}
\drawedge(n41,n42){$a$}
\drawedge(n42,n43){$b$}
\drawedge(n43,n44){$b$}

{\footnotesize
\node(n80)(20.0,-30.0){$\left[ \begin{array}{c} \# \\ \# \\ \# \\ \# \end{array}\right]$}
\node(n81)(40.0,-30.0){$\left[ \begin{array}{c} a \\ a \\ a \\ \bot \end{array}\right]$}
\node(n82)(60.0,-30.0){$\left[ \begin{array}{c} a \\ a \\ \bot \\ \bot \end{array}\right]$}
\node(n83)(80.0,-30.0){$\left[ \begin{array}{c} a \\ \bot \\ \bot \\ \bot \end{array}\right]$}
\node(n84)(100.0,-30.0){$\left[ \begin{array}{c} b \\ \bot \\ \bot \\ \bot \end{array}\right]$}
\node(n85)(120.0,-30.0){$\left[ \begin{array}{c} b \\ b \\ \bot \\ \bot \end{array}\right]$}
\node(n86)(140.0,-30.0){$\left[ \begin{array}{c} b \\ b \\ b \\ \bot \end{array}\right]$}}

\drawedge(n80,n81){$a$}
\drawedge(n81,n82){$a$}
 \drawedge(n82,n83){$a$}
 \drawedge(n83,n84){$b$}
\drawedge(n84,n85){$b$}
\drawedge(n85,n86){$b$}

\gasset{AHnb=0, dash={0.5}0}
\drawline(20, -40)(20, -43)
\drawline(80, -40)(80, -43)
\drawline(140, -40)(140, -43)

\end{picture}
\end{center}
\caption{The synchronous rational graph associated to the tiling
  system of Figure \ref{fig:ts} whose transducers are presented in
  Figure \ref{fig:ts2rat-trans}.} 
\label{fig:ts2rat-transbis}
\end{figure}

\begin{exa}
  \label{ex:ts2rat}
  Figure \ref{fig:ts2rat-trans} shows the transducers obtained using
  the previous construction on the tiling system of Figure
  \ref{fig:ts}.  They define a rational graph whose path language
  between $\sd^*$ and $b^*\bot$ is $\{ a^nb^n \mid n \geq 1 \}$.
  Figure \ref{fig:ts2rat-transbis} presents the corresponding
  synchronous graph whose vertices are the rational set of words $
  \#^{\geq 2} \; \cup \; a^+ \bot^+ \; \cup \; b^+ \bot^+ $, the set
  of initial vertices is $\#^{\geq 2}$ and the set of final vertices
  is $b^+ \bot$. Remark that in this example, the set of vertices
  accessible from the initial vertices is rational: this is not true
  in the general case.
\end{exa}

\begin{rem}
  \label{rem:bij}
  The correspondence between a tiling system $S$ and the synchronous
  graph $G$ constructed from $S$ in Proposition~\ref{prop:cs2synch} is
  tight: each picture $p$ with frontier $w$ can be mapped to a
  unique accepting path for $w$ in $G$ (and conversely).
\end{rem}

Conversely, Proposition \ref{prop:synch2cs} states that the languages
accepted by synchronous rational graphs between rational sets of
vertices can be accepted by a tiling system. To make the construction
simpler, we first prove that the sets of initial and final vertices
can be chosen over a one-letter alphabet without loss of generality. 

\begin{lem}
  \label{lem:startostar}
  For every synchronous rational graph $G$ with vertices in $\Gamma^*$
  and rational sets $I$ and $F$, one can find a synchronous rational
  graph $H$ and two symbols $i$ and $f \notin \Gamma$ such that
  $L(G,I,F) = L(H,i^*,f^*)$.
\end{lem}

\begin{proof}
  Let $G = (K_a)_{a \in \Sigma}$ be a synchronous rational graph with
  vertices in $\Gamma^*$. For $i,f$ two new distinct symbols, we
  define a new synchronous rational graph $H$ characterized by the set
  of transductions $\big(T_a = (T_I \circ K_a)\ \cup\ K_a\ \cup\ (K_a
  \circ T_F)\big)_{a \in \Sigma}$ where $T_I = \{ (i^n, u) \; | \; n
  \geq 0,\ u \in I,\ |u| = n \}$ and $T_F = \{ (v, f^n) \; | \; n \geq
  0,\ v \in F,\ |v| = n \}$. For all vertices $u \in I,\ v \in F$ we
  have $u \era{w}{G} v$ if and only if $i^{|u|} \era{w}{H} f^{|u|}$,
  i.e.  $L(G,I,F) = L(H,i^*,f^*)$.
\end{proof}

We are now able to establish the converse of Proposition
\ref{prop:cs2synch}, which states that all the languages accepted by
synchronous rational graphs between rational sets of vertices can be
accepted by a tiling system.

\begin{prop}
  \label{prop:synch2cs}
  Given a synchronous rational graph $G$ and two rational sets $I$ and
  $F$, there exists a tiling system $S$ such that ${L}(S) =
  {L}(G,I,F)$. 
\end{prop}

\begin{proof}
  Let $G = (T_a)_{a \in \Sigma}$ be a synchronous rational graph with
  vertices in $\Gamma^*$ (with $\Sigma \subseteq \Gamma$). By Lemma
  \ref{lem:startostar}, we can consider without loss of generality
  that $I = i^*$ and $F = f^*$ for some distinct letters $i$ and $f$,
  and that neither $i$ nor $f$ occurs in any vertex which is not in
  $I$ or $F$. Furthermore by Remark \ref{rem:sync-unam}, we can assume
  that $T_a$ is non-ambiguous for all $a \in \Sigma$.

  We write $Q_a$ the set of control states of $T_a$. We suppose that
  all control state sets are disjoint, and designate by $q_0^a \in
  Q_a$ the unique initial state of each transducer $T_a$, and by $Q_F$
  the set of final states of all $T_a$. 

  Let $a,b,c,d \in \Sigma$, $x,x',y,y',z,z' \in \Gamma$, and
  $p,p',q,q',r,r',s,s' \in \bigcup_{a \in \Sigma} Q_a$. We define a
  tiling system $S = (\Gamma, \Sigma, \sd, \Delta)$, where $\Delta$ is
  the set of tiles from Figure \ref{fig:synch2cs}. By construction,
  ${P}(S)$ is in exact bijection with the set of accepting paths in
  $G$ with respect to $I$ and $F$. Let $\phi$ be the function
  associating to a picture $p \in P(S)$ with columns $a_1 w_1, \ldots,
  a_n w_n$, the path $i^{|w_1|} \erb{a_1} \widetilde{w_1} \ldots
  \erb{a_n} \widetilde{w_n}$ where $\widetilde{w}$ is obtained by
  removing the control states from $w$.  By construction of $S$, the
  function $\phi$ is well defined. It is easy to check that $\phi$ is
  an onto function. As the transducers defining $G$ are non-ambiguous,
  two distinct pictures have distinct images by $\phi$ and therefore
  $\phi$ is an injection.

  \begin{figure}
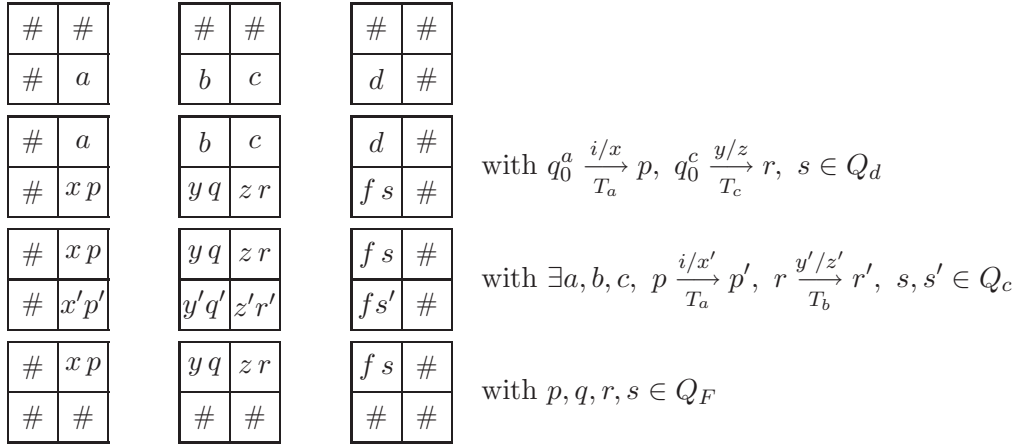

    \begin{align*}
      & \stile{\sd}{\sd}{\sd}{a} & \stile{\sd}{\sd}{b}{c} & &
      \stile{\sd}{\sd}{d}{\sd} &
      \\
      & \stile{\sd}{a}{\sd}{x\,p} & \stile{b}{c}{y\,q}{z\,r} & &
      \stile{d}{\sd}{f\,s}{\sd} & \mbox{\quad with } q_0^a
      \era{i/x}{T_a} p,\ q_0^c \era{y/z}{T_c} r,\ s \in Q_d
      \\
      & \stile{\sd}{x\,p}{\sd}{x'p'} & \stile{y\,q}{z\,r}{y'q'}{z'r'}
      & & \stile{f\,s}{\sd}{fs'}{\sd} & \mbox{\quad with } \exists
      a,b,c,\ p \era{i/x'}{T_a} p',\ r \era{y'/z'}{T_b} r',\ s,s' \in
      Q_c
      \\
      & \stile{\sd}{x\,p}{\sd}{\sd} & \stile{y\,q}{z\,r}{\sd}{\sd} & &
      \stile{f\,s}{\sd}{\sd}{\sd} & \mbox{\quad with } p,q,r,s \in Q_F
    \end{align*}
    \caption{Tiling system accepting the language of a synchronous
      graph.}
    \label{fig:synch2cs}
  \end{figure}
  
  Hence, the tiling system $(\Gamma,\Sigma,\sd,\Delta)$ exactly
  recognizes $L(G,I,F)$.
\end{proof}

\begin{rem}
  \label{rem:bij-bis}
  As in Remark \ref{rem:bij}, the set of paths in $G$ from $I$ to $F$
  and the set of pictures $P(S)$ accepted by $S$ are in bijection, and
  the length of the vertices along the path is equal to the height of
  the corresponding picture. 
\end{rem}

Putting together Propositions \ref{prop:cs2synch} and
\ref{prop:synch2cs} and Theorem \ref{thm:equiv-recogn}, we obtain the
following result concerning the path languages of synchronous rational
graphs.

\begin{thm}[\cite{Rispal02}]
  \label{th:cs-synch}
  The languages accepted by synchronous rational graphs between
  rational sets of initial and final vertices are the
  context-sensitive languages.
\end{thm}

Note that this formulation of the theorem could be made a bit more
precise by recalling that initial and final sets of vertices only of
the form $x^*$, where $x$ is a letter, are sufficient to accept all
context-sensitive languages, as stated in Lemma \ref{lem:startostar}.
By Proposition \ref{prop:rat2synch}, this implies as a corollary the
original result by Morvan and Stirling \cite{Morvan01a}.

\begin{cor}
  \label{cor:cs-rat}
  The languages accepted by rational graphs between rational sets of
  initial and final vertices are the context-sensitive languages. 
\end{cor}

If we transform a rational graph into a Turing machine by applying
successively the construction of Proposition \ref{prop:rat2synch},
Proposition \ref{prop:synch2cs} and Theorem \ref{thm:equiv-recogn}, we
obtain the same Turing machine as in \cite{Morvan01a}.

\subsection{Sequential synchronous graphs are enough}

Theorem \ref{th:cs-synch} shows that when considering rational sets of
initial and final vertices, synchronous graphs are enough to accept
all context-sensitive languages.  Interestingly, when considering
rational sets of initial and final vertices, the even more restricted
class of \emph{sequential} synchronous transducers are sufficient.

\begin{prop}
  \label{prop:cs-seq}
  The languages accepted by sequential synchronous rational graphs
  between rational sets of initial and final vertices are the
  context-sensitive languages. 
\end{prop}

\begin{proof}
  Thanks to Proposition \ref{prop:synch2cs}, it suffices to prove that
  any context sensitive language $L \subseteq \Sigma^*$ is accepted by
  a synchronous sequential rational graph. By Theorem
  \ref{thm:equiv-recogn}, we know that there exists a tiling system
  $S=(\Gamma,\Sigma,\sd,\Delta)$ such that ${L}(S)=L$.

  Let $\Lambda = \Gamma \cup \{ \sd \}$ and $[$ and $]$ be two symbols
  that do not belong to $\Lambda$. We associate to each picture $p \in
  \Lambda^{**}$ with rows $l_1, \ldots, l_n$ the word $[l_1] \ldots
  [l_n]$. We are going to define a set of sequential synchronous
  transducers that, when iterated, recognize the words corresponding
  to pictures in $P(S)$.

  First, for any finite set of tiles $\Delta$, we construct a
  transducer $T_\Delta$ which checks that a word in $([\Lambda^{\geq
    3}])^{\geq 2}$ represents a picture with tiles in $\Delta$. The
  checking is done column by column, and we introduce marked letters
  to keep track of the column being checked. Let
  $\widetilde{{\Lambda}}$ be a finite alphabet in bijection with but
  disjoint from $\Lambda$.  For all $x \in \Lambda$ we write
  $\widetilde{{x}} \in \widetilde{{\Lambda}}$ the marked version of
  $x$.  For every word $w = u \widetilde{{x}} v \in \Lambda^*
  \widetilde{{\Lambda}} \Lambda^*$, we write $\pi(w)$ the word $u x v
  \in \Lambda^*$ and $\rho(w)=|u|+1$ designates the position of the
  marked letter in the word.

  We consider words in $[\Lambda^* \widetilde{{\Lambda}}
  \Lambda^*]^{\geq 2}$.  Let $\mathrm{Shift}$ be the relation that
  shifts all marks in a word one letter to the right. More precisely,
  $\mathrm{Shift}$ satisfies $Dom(\mathrm{Shift})=([\Lambda^*
  \widetilde{{\Lambda}} \Lambda^+])^{\geq 2}$, and
  $\mathrm{Shift}([w_1] \ldots [w_n]) = [w'_1] \ldots [w'_n]$ with
  $\pi(w'_i)=\pi(w_i)$ and $\rho(w'_i) = \rho(w_i)+1$ for all $i \in
  [1,n]$. The rational relation $\mathrm{Shift}$ can be realized by a
  synchronous sequential transducer $T_\mathrm{Sh}$. Consider the
  following rational language:
  \[
  R_\Delta = \left\{\, [w_1 x_1 \widetilde{{y_1}} w'_1] \ldots [w_n
    x_n \widetilde{{y_n}} w'_n] \mid n\geq 2\; \text{and} \; \forall i
    \in [2,n],\
    \begin{array}{|@{}c@{}|@{}c@{}|}
      \hline 
      \minipage[c][\cellsize][c]{\cellsize+2pt} 
      \centering{$x_{i-1}$} \endminipage & 
      \minipage[c][\cellsize][c]{\cellsize+2pt} 
      \centering{$y_{i-1}$} \endminipage
      \\
      \hline
      \cell{$x_{i}$} & \cell{$y_{i}$}
      \\
      \hline
    \end{array} \in \Delta\, \right\}. 
  \]
  The transducer $T_\Delta$ obtained by restricting $T_\mathrm{Sh}$ to
  the domain $R_\Delta$ is both synchronous and sequential. For all
  $w=[w_1] \ldots [w_n] \in ([\Lambda \widetilde{{\Lambda}}
  \Lambda^*])^{\geq 2}$, if $w'= T_\Delta^N(w)$ then $w'=[w'_1] \ldots
  [w'_n]$ with $\pi(w_i)=\pi(w'_i)$ and $\rho(w_i')=N+2$ for all $i
  \in [1,n]$. Let $r_i$ be the word containing the $N+1$ first letters
  of $w'_i$, a straightforward induction on $N$ proves that the
  picture $p$ formed of the rows $r_1, \ldots, r_n$ only has tiles in
  $\Delta$. In particular, $T_\Delta^N(w)$ belongs to $([\Lambda^*
  \widetilde{{\Lambda}}])^* \cap R_\Delta$ if and only if $\pi(w)$
  represent a picture $p$ of width $N+2$ such that $T(p) \subseteq
  \Delta$.

  We now define more precisely the sequential rational graph $G =
  (T_a)_{a \in \Sigma}$ accepting $L$. For all $a \in \Sigma$, the
  transducer $T_a$ is obtained by restricting the domain of $T_\Delta$
  to the set of words representing pictures whose marked symbol on the
  second row is $a$, i.e. to the set $[(\Lambda \cup
  \widetilde{{\Lambda}})^*][(\Lambda^* \widetilde{{a}}
  \Lambda^*][(\Lambda \cup \widetilde{{\Lambda}})^*]^*$. $T_a$ can be
  chosen synchronous and sequential.  The set of initial vertices $I$
  is $[\sd \widetilde{{\sd}} \sd^*] ([\sd \widetilde{{\Gamma}}
  \Gamma^* \sd])^* [\sd \widetilde{{\sd}} \sd^*]$ and the set of final
  vertices $F$ is $[\sd^* \widetilde{{\sd}}] ([\sd \Gamma^*
  \widetilde{{\sd}}])^* [\sd^*\widetilde{{\sd}}]$.
\end{proof}

\begin{figure}
  \begin{center}
    \setlength{\arraycolsep}{1pt}
    \renewcommand{\arraystretch}{0.5}
    \begin{picture}(150,23)(-12,-12)
      \gasset{Nadjust=wh,Nadjustdist=1,Nframe=n}
      
      \node(a)(0,0){$
        \begin{array}{cccccc}
          \sd & \widetilde{{\sd}} & \sd & \sd & \sd & \sd \\
          \sd & \widetilde{{a}}  & a   & b   & b   & \sd \\
          \sd & \widetilde{{a}}  & \bot & \bot & b   & \sd \\
          \sd & \widetilde{{\bot}} & \bot & \bot & \bot & \sd \\
          \sd & \widetilde{{\sd}} & \sd & \sd & \sd & \sd \\
        \end{array}$}
      \node(b)(32,0){$
        \begin{array}{cccccc}
          \sd & \sd & \widetilde{{\sd}} & \sd & \sd & \sd \\
          \sd & a   & \widetilde{{a}}  & b   & b   & \sd \\
          \sd & a   & \widetilde{{\bot}} & \bot & b   & \sd \\
          \sd & \bot & \widetilde{{\bot}} & \bot & \bot & \sd \\
          \sd & \sd & \widetilde{{\sd}} & \sd & \sd & \sd \\
        \end{array}$}
      \node(c)(64,0){$
        \begin{array}{cccccc}
          \sd & \sd & \sd & \widetilde{{\sd}} & \sd & \sd \\
          \sd & a   & a   & \widetilde{{b}}  & b   & \sd \\
          \sd & a   & \bot & \widetilde{{\bot}} & b   & \sd \\
          \sd & \bot & \bot & \widetilde{{\bot}} & \bot & \sd \\
          \sd & \sd & \sd & \widetilde{{\sd}} & \sd & \sd \\
        \end{array}$}
      \node(d)(96,0){$
        \begin{array}{cccccc}
          \sd & \sd & \sd & \sd & \widetilde{{\sd}} & \sd \\
          \sd & a   & a   & b   & \widetilde{{b}} & \sd \\
          \sd & a   & \bot & \bot & \widetilde{{b}} & \sd \\
          \sd & \bot & \bot & \bot & \widetilde{{\bot}} & \sd \\
          \sd & \sd & \sd & \sd & \widetilde{{\sd}} & \sd \\
        \end{array}$}
      \node(e)(128,0){$
        \begin{array}{cccccc}
          \sd & \sd & \sd & \sd & \sd & \widetilde{{\sd}} \\
          \sd & a   & a   & b   & b   & \widetilde{{\sd}} \\
          \sd & a   & \bot & \bot & b   & \widetilde{{\sd}} \\
          \sd & \bot & \bot & \bot & \bot & \widetilde{{\sd}} \\
          \sd & \sd & \sd & \sd & \sd & \widetilde{{\sd}} \\
        \end{array}$}
      
      \drawedge(a,b){$a$}
      \drawedge(b,c){$a$}
      \drawedge(c,d){$b$}
      \drawedge(d,e){$b$}
    \end{picture}
  \end{center}
  \caption{Connected component of a sequential synchronous graph
    accepting $\{ a^nb^n | n \geq 1 \}$.}
  \label{fig:sequ-rat}
\end{figure}
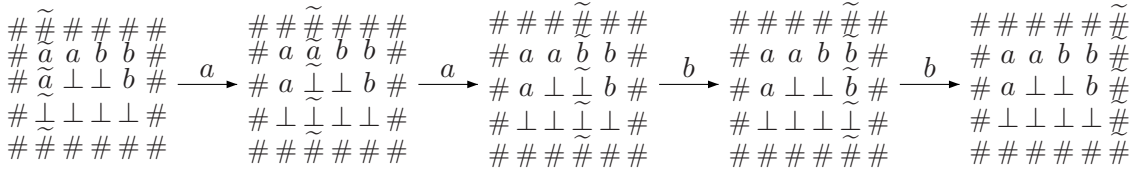

\begin{exa}
  Figure \ref{fig:sequ-rat} shows a part of the result of the previous
  construction when applied to the language $\{ a^nb^n \mid n \geq 1
  \}$ as recognized by the tiling system of Figure \ref{fig:ts}. Each
  vertex is represented by the corresponding picture, instead of the
  word coding for it. Also, only one connected component of the graph
  is shown. The other connected components all have the same linear
  structure: the degree of the graph is bounded by 1. The leftmost
  vertex belongs to the set $I$, and the rightmost to the set $F$,
  hence the word $a^2b^2$ is accepted.
\end{exa}

\begin{rem}
  In the case of synchronized transducers, it has been shown in Lemma
  \ref{lem:startostar} that $I$ could be taken over
  a one letter alphabet without loss of generality.  This does not
  seems to hold for sequential transducers as the proof we present
  relies on the expressiveness of the initial set of vertices. 
  In fact, as shown in Proposition \ref{prop:quitue}, the languages 
  recognized by sequential synchronous graph from $i^*$ are deterministic
  context-sensitive languages. 
\end{rem}

\section{Rational graphs seen as automata}
\label{sect:automata}

The structure of the graphs obtained in the previous section
(propositions \ref{prop:cs2synch} and \ref{prop:cs-seq}) is very poor.
Synchronous graphs are by definition composed of a possibly infinite
set of \emph{finite} connected components.  In the case of
Proposition~\ref{prop:cs-seq}, we obtain an even more restricted
family of graphs since both their in-degree and out-degree is bounded
by 1. However, when considering accepted languages from a possibly
infinite rational set of vertices, even this extremely restricted
family accepts the same languages as the most general rational graphs,
namely all context-sensitive languages. This is why, in order to
compare the expressiveness of the different sub-families of rational
graphs and to obtain graphs with richer structures, we need to impose
structural restrictions.

We first consider graphs with a single initial vertex, but this
restriction alone is not enough. In fact, both synchronized and
rational graphs with a rational set of initial vertices accept the
same languages as their counterparts with a single initial vertex.

\begin{lem} 
  \label{lem:one-point} 
  For every rational graph (resp.  synchronized graph) $G$ and for
  every pair of rational sets $I$ and $F$, there exists a rational
  graph (resp. a synchronized graph) $G'$, a vertex $i$ and a rational
  set $F'$ such that $L(G,I,F) = L(G',\{i\},F')$.
\end{lem}

\begin{proof}
  Let $G = (T_a)_{a \in \Sigma}$ be a rational graph with vertices in
  $\Gamma^*$ and let $i$ be a symbol which does not belong to $\Gamma$
  and $\Gamma'=\Gamma \cup \{i\}$. For all $a \in \Sigma$, let $T'_a$
  be a transducer recognizing the rational relation $T_a \cup \{ (i,w)
  \mid w \in T_a(I)\}$. Remark that if $T_a$ is synchronized then
  $T'_a$ can also be chosen synchronized.  If $\eps \not\in L(G,I,F)$
  then $F'=F$ else $F'=F \cup \{i\}$. It is straightforward to show
  that $L(G,I,F) = L(G',\{i\},F')$.
\end{proof}

It follows from Proposition~\ref{prop:cs2synch} and Lemma
\ref{lem:one-point} that the synchronized rational graphs with one
initial vertex accept the context-sensitive languages \cite{Rispal02}.

\begin{rem} 
  \label{rem:one-point-synch} 
  It is fairly obvious that this result does not hold for synchronous
  graphs: indeed, the restriction of a synchronous rational graph to
  the vertices reachable from a single vertex is finite. Hence, the
  languages of synchronous graphs from a single vertex are rational.
  Similarly, as any rational language is accepted by a deterministic
  finite graph, it can also be accepted by a sequential synchronous
  graph with a single initial vertex.
\end{rem}

Note that the construction of Lemma \ref{lem:one-point} relies on
infinite out-degree to transform a synchronous graph with a rational
set of initial vertices into a rational one with a single initial
vertex. In order to obtain more satisfactory notions of infinite
automata, we now restrict our attention to graphs of finite out-degree
with a single initial vertex.

\subsection{Rational graphs of finite out-degree with one initial
  vertex.}

We present a syntactical transformation of a synchronous rational
graph with a rational set of initial vertices into a rational graph of
finite out-degree with a unique initial vertex accepting the same
language.

The construction relies on the fact that for a synchronous graphs to
recognize a word of length $n>0$, it is only necessary to consider
vertices whose length is smaller than $c^n$ (where $c$ is a constant
depending only on the graph). We first establish a similar result for
tiling systems and conclude using the close correspondence between
synchronous graphs and tiling systems established in
Proposition~\ref{prop:synch2cs}.

\begin{lem}
  \label{lem:bound-ts}
  For any tiling system ${S}=(\Gamma,\Sigma,\#,\Delta)$,
  if $p \in {P}({S})$ then there exists a
  $(n,m)$-picture $p'$ such that $\fr{p}=\fr{p'}$ and $n \leq |\Gamma|^m$.
\end{lem}

\begin{proof}
Let $p'$ be a $(n,m)$-picture with $n > |\Gamma|^m$, and suppose that
$p'$ is the smallest picture in ${P}({S})$ with frontier $\fr{p}$. Let
$l_1,\ldots,l_n$ be the rows of $p'$. As $n > |\Gamma|^m$ then there
exists $j>i \geq 1$ such that $l_i=l_j$. Let $p''$ be the picture with
rows $l_1, \ldots, l_i,l_{j+1},\ldots, l_n$.  It is easy to check that
$T({p''}_{\td}) \subset T({p'}_{\td})$, we have that $p'' \in
{P}({S})$ and as $p''$ has a smaller height than $p'$ but the same
frontier, we obtain a contradiction. 
\end{proof}

We know from Remark~\ref{rem:bij-bis} that for every synchronous
rational graph $G=(T_a)_{a \in \Sigma}$ and two rational sets $I$ and
$F$, there exists a tiling system ${S}$ such that $i \era{w}{G} f$
with $i \in I$ and $f \in F$ if and only if there exists $p \in K$
such that $\fr{p}=w$ and $p$ has height $|i|=|f|$. Hence, as a direct
consequence of Lemma \ref{lem:bound-ts}, one gets:

\begin{lem}
  \label{lem:synch-bound}
  For every synchronous rational graph $G$ and rational sets $I$ and
  $F$, there exists $k \geq 1$ such that:
  \[
  \forall w \in L(G,I,F), \exists i \in I, f \in F \text{ such that }
  i \era{w}{G} f \text{ and } |i| = |f| \leq k^{|w|}.
  \]
\end{lem}

We can now construct of a rational graph of finite out-degree
accepting from a single vertex the same language as a synchronous
graph with a rational set of initial vertices.

\begin{prop}
  \label{prop:synch2ratfd}
  For every synchronous rational graph $G$ and rational sets $I$ and
  $F$ such that $I \cap F = \emptyset$, there is a rational graph $H$
  of finite out-degree and a vertex $i$ such that $L(G,I,F) =
  L(H,\{i\},F)$.
\end{prop}

\begin{figure}
    \begin{center}
      \input{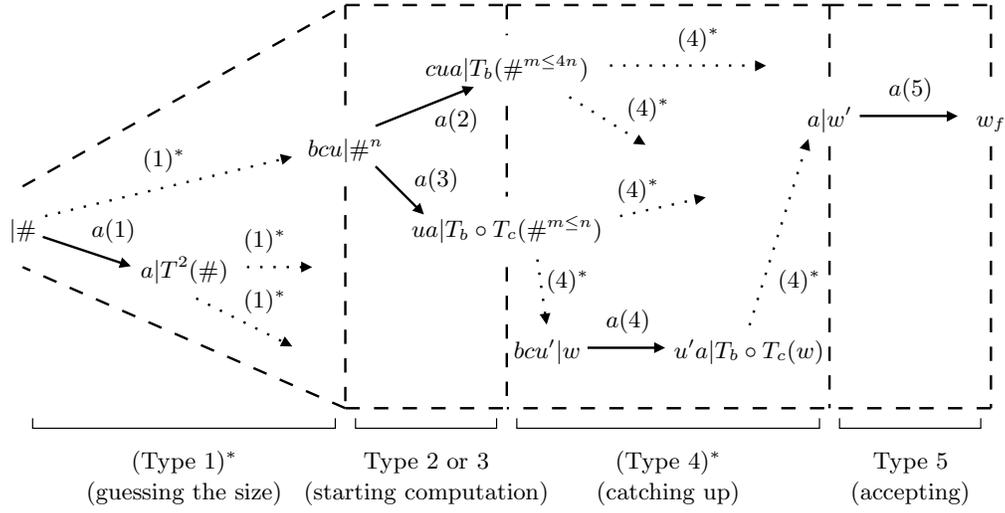}
    \end{center}
    \caption{Schema of the construction in Proposition
      \ref{prop:synch2ratfd}.}
    \label{fig:ratfd}
  \end{figure}

\begin{proof}
  According to Lemma \ref{lem:startostar}, there exists a synchronous
  rational graph $R$ described by a set of transducers $(T_a)_{a \in
    \Sigma}$ over $\Gamma^*$ such that $L(G,I,F) = L(R,\sd^*,F)$. Note
  that for all $w \in \sd^*$ and $w' \in \Gamma^*$, if $w \era{}{R}
  w'$ then $w'$ does not contain $\sd$.  We define a graph $H$
  such that $L(G,I,F) = L(H,\{i\},F)$ for some vertex $i$ of $H$. Let
  $k$ be the constant involved in Lemma \ref{lem:synch-bound}, $T$ and
  $T'$ two transducers realizing the rational relations $\left\{
    (\sd^n,\sd^{kn}) \;|\; n \in \mathbb{N} \right\}$ and $\left\{
    \sd^n,\sd^m \; | \; m \in [1,n] \right\}$ respectively.  For all
  $a,b,c \in \Sigma$ and $u \in \Sigma^*$, $H$ has edges:
  \begin{align*}
    \forall n \in \mathbb{N}, & \quad &u & | \sd^n & \erb{a} & & ua
    &|\, T \circ T(\sd^n) & & \text{(Type 1)}
    \\
    \forall n \in \mathbb{N}, & & bu & | \sd^n & \erb{a} & & ua & |\,
    T \circ T' \circ T_b(\sd^{n}) & & \text{(Type 2)}
    \\
    \forall n \in \mathbb{N}, & & bcu & | \sd^n & \erb{a} & & ua & |\,
    T' \circ T_b \circ T_c(\sd^{n}) & & \text{(Type 3)}
    \\
    \forall w \in \left(\Gamma \setminus \{\sd\} \right)^*, & & bcu &
    | w & \erb{a} & & ua & |\, T_b \circ T_c(w) & & \text{(Type 4)}
    \\
    \forall w \in \left(\Gamma \setminus \{\sd\} \right)^*, & & b & |
    w & \erb{a} & & & T_b \circ T_a(w) & & \text{(Type 5)}
    \\
    & & & | \sd & \erb{a} & & & T \circ T' \circ T_a(\sd) & &
    \text{(Type 6)}
  \end{align*}
  The graph $H$ is clearly rational and of finite out-degree. We take
  $i=|\sd$ as initial vertex.  Remark that in $H$ an edge of type 2 or
  3 cannot be followed by edges of type 1, 2 or 3, and at most one
  edge of type 2 or 3 and of type 5 or 6 can be applied. Moreover, an
  edge of type 1 increases the length of the left part of the word by
  one, and an edge of type 4 decreases it by one. Also, in any
  accepting path, the last edge is of type 5 or 6. Figure
  \ref{fig:ratfd} illustrates the structure of the obtained graph. It
  is technical but straightforward to show a correspondence between
  accepting paths in $H$ and $R$, and to conclude that $L(R,\sd^*,F) =
  L(H,\{i\},F)$.
\end{proof}

From Proposition \ref{prop:rat2synch} and
Proposition~\ref{prop:synch2ratfd}, we deduce that the rational graphs
of finite out-degree with one initial vertex accept all
context-sensitive languages. This result was proved in
\cite{Morvan01a} using the Penttonen normal form of context-sensitive
grammars \cite{Penttonen74}.

\begin{thm}
  \label{th:cs-ratfd}
  The path languages of rational graphs of finite out-degree from a
  unique initial vertex to a rational set of final vertices are the
  context-sensitive languages.
\end{thm}

\subsection{Synchronized graphs of finite out-degree with one initial
  vertex}

We now consider the languages of synchronized graphs of finite
out-degree with one initial vertex. First, we characterize them as the
languages recognized by tiling systems with square pictures (i.e.  for
which there exists $c \in \mathbb{N}$ such that for every word $w \in
{L}(S)$, there exists a $(n,m)$-picture in $P(S)$ with $n \leq c m$
and with frontier $w$). A slight adaptation of the construction of
Proposition~\ref{prop:synch2ratfd} gives the first inclusion. 

\begin{prop} 
  \label{prop:synchfd2squarets} 
  Let $S=(\Gamma,\Sigma,\#,\Delta)$ be a tiling system with square
  pictures. There exists a synchronized rational graph of finite
  degree accepting ${L}(S)$ from one initial vertex.
\end{prop}
 
\begin{proof}
  Let $G=(T_a)_{a \in \Sigma}$ be the synchronized graph obtained from
  $S$ in Proposition~\ref{prop:cs2synch}.  In the construction from
  the proof of Proposition~\ref{prop:synch2ratfd}, if we replace the
  transducer $T$ by a transducer $S$ realizing the synchronized
  relation $\{ (\#^n,\#^{n+c}) \;|\; n \in \mathbb{N} \}$, we obtain a
  synchronized graph $H$, a vertex $i$ and a set $F$ such that
  ${L}(H,i,F) = {L}({S})$.
\end{proof}
 
Before proceeding with the converse, we state a result similar to
Lemma~\ref{lem:synch-bound} for synchronized graphs \emph{of finite
  out-degree} that states that when recognizing a word $w$ \emph{from
  a unique initial vertex} $i$, the vertices involved have a length at
most linear in the size of $w$.
 
\begin{lem}
  \label{lem:synchronized-bound}
  For any synchronized rational graph $G$ of finite out-degree with
  vertices in $\Gamma^*$ and for every vertex $i$, there exists a
  constant $k$ such that for all $w$ in $L(G,\{i\},F)$, there exists a
  path from $i$ to some $f \in F$, labeled by $w$, and with vertices
  of size at most $k \cdot |w|$.
\end{lem}

\begin{proof}
  It follows from the definition of synchronized transducers that for
  every synchronized transducer of finite out-degree there exists $c
  \in \mathbb{N}$ such that $(x,y) \in T$ implies that $|x| \leq |y| +
  c$ (see \cite{Sakarovitch03} for a proof of this result). We take
  $k$ to be the maximum over the set of transducers defining $G$ of
  these constants.  The result follows by a straightforward induction
  on the size of $w$.
\end{proof}

The converse inclusion is obtained by remarking that the composition
of the construction of Proposition~\ref{prop:rat2synch} and
Proposition~\ref{prop:synch2cs} gives a tiling system with square
pictures when applied to a synchronized graph of finite out-degree.

\begin{prop} 
  \label{prop:squarets2synchfd} 
  Let $G=(T_a)_{a \in \Sigma}$ be a synchronized graph of finite
  out-degree. For every initial vertex $i$ and set of final vertices
  $F$, there exists a tiling system ${S}$ with square pictures such
  that ${L}({S})= {L}(G,\{i\},F)$.
\end{prop}
 
\begin{proof}
  Let $G'$, $I'$ and $F'$ be the synchronous graph and the rational
  set of initial and final vertices obtained by applying the
  constructions of Proposition~\ref{prop:rat2synch} to $G$, $\{i\}$
  and $F$.  It is easy to show that for every word $w \in
  {L}(G',I',F')$, there exists $i' \in I'$ and $f' \in F'$ such that
  $i' \eRb{w} f'$ with $|i'|=|f'| \leq k |w|$ where $k$ is the
  constant of Lemma~\ref{lem:synchronized-bound} for $G$. We conclude
  by Proposition~\ref{prop:synch2cs}, that states the existence of a
  tiling system $S$ such that ${L}(S)= {L}(G',I',F')$. By Remark
  \ref{rem:bij}, $S$ is a tiling system with square pictures.
\end{proof}

Putting together Proposition~\ref{prop:squarets2synchfd} and
Proposition~\ref{prop:synchfd2squarets} and with the use of the
simulation result from Theorem~\ref{thm:equiv-recogn}, we obtain the
following theorem. 

\begin{thm}
  \label{thm:carac-sdb}
  The languages accepted by synchronized graphs of finite out-degree
  from a unique vertex to a rational set of vertices are the
  context-sensitive languages recognized by non-deterministic linearly
  bounded machines with a linear number of head reversals. 
\end{thm}

We conjecture that this class is strictly contained in the
context-sensitive languages. However, few separation results exist for
complexity classes defined by time and space restrictions (see for
example \cite{Melkebeek04}). In particular, the diagonalization
techniques (see \cite{Fortnow00}) used to prove that the polynomial
time hierarchy (with no space restriction) is strict do not apply for
lack of a suitable notion of \emph{universal} LBM.

\subsection{Bounding the out-degree}

It is natural to wonder if the rational graphs still accept the
context-sensitive languages when considering bounded out-degree. This
is a difficult question, to which we only provide here a partial
answer concerning the synchronized graphs of bounded out-degree. 

It follows from Lemma~\ref{lem:synchronized-bound} that the vertices
used to accept a word $w$ in a synchronized rational graph have a
length at most linear in the length of $w$ and therefore, can be
stored on the tape of a LBM. Moreover if the graph is deterministic,
we can construct a deterministic LBM accepting its language. 

\begin{prop}
  \label{prop:det-bound-synch}
  The language accepted by a deterministic synchronized graph from a
  unique initial vertex is deterministic context-sensitive. 
\end{prop}

\begin{proof}
  Let $G = (T_a)_{a \in \Sigma}$ be a deterministic synchronized graph
  over $\Gamma$, $i$ a vertex and $F$ a rational set of vertices.  We
  define a deterministic LBM $M$ accepting ${L}(G,\{i\},F)$.  When
  accepting $w = a_1 \ldots a_{|w|}$, $M$ starts by writing $i$ on its
  tape. It successively applies $T_{a_1}$, \ldots, $T_{a_{n-1}}$ and
  $T_{a_n}$ to $i$.  If the image of the current tape content by one
  of these transducers is not defined, the machine rejects.
  Otherwise, it checks whether the last tape content represents a
  vertex which belongs to $F$.

  We now detail how the machine $M$ can apply one of the transducers
  $T$ of $G$ to a word $x$ in a deterministic manner. As $T$ has a
  finite image, we can assume without loss of generality that $T =
  (\Gamma, Q, i, F, \delta)$ is in real-time normal form: $\delta
  \subset Q \times \Gamma \times \Gamma^*\times Q$ (see for instance
  \cite{Berstel79} for a presentation of this result). The machine
  enumerates all paths in $T$ of length less than $c |x|$ in the
  lexicographic order where $c$ is the constant associated to $G$ in
  Lemma \ref{lem:synchronized-bound}. For each such path $\rho$, it
  checks if it is an accepting path for input $x$, and in that case
  replaces $x$ by the output of $\rho$.

  The space used by $M$ when starting with a word $w$ is bounded by
  $(2c+1)|w|$. Moreover if $M$ accepts $w$, then there exists a path
  from $i$ to a vertex $F$ in $G$ labeled by $w$. Conversely, if $w$
  belongs to ${L}(G,\{i\},F)$ then by
  Lemma~\ref{lem:synchronized-bound}, there exists a path in $G$ from
  $i$ to $F$ with vertices of length at most $c |w|$ and by
  construction $M$ accepts $w$. Hence, $M$ is a deterministic linearly
  bounded Turing machine accepting ${L}(G,\{i\},F)$.
\end{proof}

\begin{rem}
  \label{rem:det}
  The result of Proposition~\ref{prop:det-bound-synch} extends to any
  deterministic rational graph satisfying the property expressed by
  Lemma~\ref{lem:synchronized-bound}. 
\end{rem}

The previous result can be extended to synchronized graphs of bounded
out-degree thanks to a uniformization result by Weber. First observe
that a rational graph is of out-degree bounded by some constant $k$ if
and only if it is defined by transducers which associate at most $k$
distinct images to any input word. The relations realized by these
transducers are called $k$-valued rational relations. 

\begin{prop}[\cite{Weber96}]
  \label{prop:unif}
  For any $k$-valued rational relation $R$, there exist $k$ functional
  rational relations $F_1,\ldots,F_k$ such that $R = \bigcup_{i \in
    [1,k]} F_i$. 
\end{prop}

Note that even if $R$ is a synchronized relation, the $F_i$'s are not
necessarily synchronized. However, they still satisfy the inequality
$|y| \leq |x| + c$ for all $(x,y)\in F_i$. 

To any synchronized graph $G$ with an out-degree bounded by $k$
defined by a set of transducers $(T_a)_{a \in \Sigma}$, we associate
the deterministic rational graph $H$ defined by $(F_{a_i})_{a \in
  \Sigma, i \in [1,k]}$ where for all $a \in \Sigma$, $(F_{a_i})_{i
  \in [1,k]}$ is the set of rational functions associated to $T_a$ by
Proposition~\ref{prop:unif}. According to
Proposition~\ref{prop:det-bound-synch} and to Remark~\ref{rem:det},
${L}(H,\{i\},F)$ is a deterministic context-sensitive
language. Let $\pi$ be the alphabetical projection defined by
$\pi(a_i)=a$ for all $a \in \Sigma$ and $i \in [1,k]$, it is
straightforward to establish that $\pi\left({L}(H,\{i\},F)
\right) = {L}(G,\{i\},F)$. As deterministic context-sensitive
languages are closed under alphabetical projections,
${L}(G,\{i\},F)$ is a deterministic context-sensitive
language. 

\begin{thm}
\label{thm:bound-synch}
The language accepted by a synchronized graph of bounded out-degree
from a unique initial vertex is deterministic context-sensitive. 
\end{thm}

The converse result is not clear, for reasons similar to those
presented in the previous section for synchronized graphs of finite
degree. A precise characterization of the family of languages accepted
by synchronized rational graphs of bounded degree would be interesting.

\section{Notions of determinism}
\label{sect:determinism}

In this last part of the section on rational graphs, we investigate
families of graphs which accept the deterministic context-sensitive
languages.  First of all, we examine the family yielded by the
previous constructions when applied to deterministic languages. Then,
we propose a global property over sets of transducers characterizing a
sub-family of rational graphs whose languages are precisely the
deterministic context-sensitive languages.

\subsection{Unambiguous context-sensitive languages}

When applying the construction of Proposition \ref{prop:cs2synch} to a
deterministic tiling system ${S}$, one obtains a synchronous rational
graph $G$ (which is non-deterministic in general) and two rational
sets of vertices $I$ and $F$ such that $L(G,I,F) = L({S})$, with the
particularity that for every word $w$ in $L$, there is \emph{exactly
  one} path labeled by $w$ leading from some vertex in $I$ to a vertex
in $F$: $G$ is \emph{unambiguous} with respect to $I$ and $F$.
However, the converse is not granted: given a graph $G$ and two
rational sets $I$ and $F$ such that $G$ is unambiguous with respect to
$I$ and $F$, we cannot ensure that $L(G,I,F)$ is a deterministic
context-sensitive language. Rather, the obtained languages can be
accepted by unambiguous linearly bounded machines. This class of
languages is called $\mathrm{USPACE}(n)$, and it is not known whether
it coincides with either the context-sensitive or deterministic
context-sensitive languages.

\begin{thm}
\label{thm:cs-unamb}
  Let $L$ be a language, the following properties are equivalent:
  \begin{enumerate}
  \item $L$ is an unambiguous context-sensitive language. 
  \item There exist a rational graph $G$ with unambiguous transducers and
    two rational sets $I$ and $F$ with respect to which $G$ is
    unambiguous, such that $L = {L}(G, I, F)$.
  \end{enumerate}
\end{thm}

This result only holds if one considers unambiguous transducers, i.e.
transducers in which there is at most one accepting path per pair of
words. The reason is that ambiguity in the transducers would induce
ambiguity in the machine. However, since synchronized transducers can
be made unambiguous (Cf. Remark \ref{rem:sync-unam}), we can drop this
requirement in the case of synchronized graphs. Note that the
unambiguity of a rational or synchronized graphs with respect to
rational sets of vertices is undecidable. However, since any rational
function can be realized by an unambiguous transducer
\cite{Kobayashi69,Sakarovitch03}, the language of any deterministic
rational graph is, by to Theorem~\ref{thm:cs-unamb}, unam\-bi\-guous.

\begin{cor}
  The languages of deterministic rational graphs from an initial
  vertex $i$ to a rational set $F$ of vertices are unambiguous
  context-sensitive languages.
\end{cor}

\subsection{Globally deterministic sets of transducers}

We just saw an attempt at characterizing natural families of graphs
whose languages are the deterministic context-sensitive languages,
which was based on a restriction of previous constructions to the
deterministic case, but failed to meet its objective because of a
slight nuance between the notions of determinism and unambiguity for
tiling systems.

First, we naturally consider the class of sequential synchronous
automata with an initial set of the form $\{a\}^*$, where $a$ is a
letter of the vertex alphabet (in other words, a given initial vertex
does not code for any information besides its length). It is easy to
check that when applying the construction of Proposition
\ref{prop:synch2cs} to one of these automata, we obtain a
deterministic tiling system.

\begin{prop}
  \label{prop:quitue}
  The languages of sequential synchronous graphs from $\{a\}^*$ are
  deterministic context-sensitive languages.
\end{prop}

The converse result seems difficult to prove due to the local
nature of the determinism involved in this class. Hence, we consider a
\emph{global} property of the set of transducers characterizing a
rational graph, so as to ensure that each accepting path corresponds
to the run of a deterministic linearly bounded machine on the
corresponding input, or equivalently that each accepting path
corresponds to a picture recognized by a deterministic tiling system
and whose upper frontier is the path label under consideration. 

For any rational language $L$, we write $T_L$ the minimal synchronous
transducer recognizing the identity relation over $L$. 

\begin{defi}
  Let $T$ be a set of synchronous transducers over $\Gamma$. We say
  $T$ is \emph{globally deterministic} with respect to two rational
  languages $I$ and $F \subseteq \Gamma^*$ if all transducers in $T$
  are deterministic\footnotemark\ and for every pair of transducers $T_1
  \in T \cup \{T_I\}$ and $T_2 \in T \cup \{T_F\}$, and every pair of
  control states $q_1 \in Q_{T_1}$ and $q_2 \in Q_{T_2}$, there is at
  most one $b$ such that
  \[
  q_1 \era{a/b}{T_1} q'_1 \land q_2 \era{b/c}{T_2} q'_2 \text{ for
    some } a,c \in \Gamma,\ q'_1 \in Q_{T_1},\ q'_2\in Q_{T_2}.
  \]
\end{defi}
\footnotetext{i.e. whenever $q \era{a/b}{} q'$ and $q \era{c/d}{} q''$
  with $q' \not= q''$, it implies $(a,b) \neq (c,d)$.}

Intuitively, this condition states that, whenever a part of the output
of one transducer can be read as input by a second transducer, there
is only one way to add a letter to this word such that it is still
compatible with both transducers. This property of sets of transducers
is trivially decidable, since it is sufficient to check the above
condition for every pair of control states of transducers in $(T \cup
\{T_I\}) \times (T \cup \{T_F\})$. This allows us to capture a
sub-family of rational graphs whose languages are the deterministic
context-sensitive languages. 

\begin{thm}
  \label{thm:glob-det}
  Let $L$ be a language, the following two properties are equivalent:
  \begin{enumerate}
  \item $L$ is a deterministic context-sensitive language. 
  \item There is a synchronous rational graph $G$ and two rational
    sets $I$ and $F$ such that $L = {L}(G,I,F)$ and $G$ is globally
    deterministic between $I$ and $F$. 
  \end{enumerate}
\end{thm}

\begin{proof}
  Let $G=(T_a)_{a \in \Sigma}$ be a synchronous rational graph which
  is globally deterministic between $I$ and $F$.  The graph
  $H=(T'_a)_{a \in \Sigma}$ obtained by applying Lemma
  \ref{lem:startostar} to $G$ is such that $L(H,i^*,f^*)=L(G,I,F)$.
  Moreover, $H$ is globally deterministic between $i^*$ and $f^*$.

  We will show that the construction of Proposition
  \ref{prop:synch2cs}, when applied to a rational graph $H$ between
  $i^*$ and $f^*$ yields a deterministic tiling system. Suppose that
  this is not the case. Then, by definition of a non-deterministic
  tiling system, there must be words $u$, $v_1$ and $v_2$ with $v_1
  \neq v_2$ such that the two-rows pictures $p_1$ and $p_2$ with first
  row $\sd u \sd$ and second row $\sd v_1 \sd$ and $\sd v_2 \sd$
  respectively only have tiles in $\Delta$. Since $v_1 \neq v_2$, let
  $i$ be the smallest index such that $v_1(i) \neq v_2(i)$. Let
  $v_1(i) = xp$, $v_2(i) = x'p'$.

  By the construction of Prop.~\ref{prop:synch2cs}, there are two
  transducers $T_a$ and $T_b$ such that
  \[
  q_a \era{y/x}{T_a} p \land q_b \era{x/z}{T_b} q'_b \land q_a
  \era{y/x'}{T_a} p' \land q_b \era{x'/z'}{T_b} q''_b
  \]
  for some symbols $y,y',z,z' \in \Gamma$ and control states $q_a,
  q_b, q'_b$ and $q''_b$.  As $T_a$ is deterministic, if $x$ is equal
  to $x'$, then $p=p'$ and $v_1(i)=v_2(i)$. Hence $x \not= x'$, and
  the above relations contradicts the global determinacy of $H$.

  To prove the converse, we introduce yet another family of acceptors
  for context-sensitive languages, namely cellular automata. A
  cellular automaton is a tuple $(\Gamma,\Sigma,F,[,],\delta)$ where
  $\Gamma$ and $\Sigma \subseteq \Gamma$ are the work and input
  alphabets, $F \subseteq \Gamma$, $[,] \not\in \Gamma$ and $\delta$
  is a set of 4-tuples over $\Gamma$ called transition rules. These
  rules induce a transition relation over words of the form $[u] \in
  [\Gamma^*]$: $c' = [v]$ is a successor of $c = [u]$ if $|c| = |c'| =
  n$ and for all $i \in [2,n-1]$, $(c(i-1), c(i), c(i+1), c'(i)) \in
  \delta$. A word $w$ is accepted if, starting from $[w]$ one can
  derive a word $[u]$ with $u \in F^*$. An cellular automaton is
  deterministic if for all $A,B,C$ there is at most one $D$ such that
  $(A,B,C,D) \in \delta$.  Equivalence of (deterministic) cellular
  automata with (deterministic) LBMs or tiling systems is common
  knowledge.

  Let $L$ be any deterministic context-sensitive language, there
  exists a deterministic cellular automaton ${C} = (\Gamma, \Sigma,
  \bot, \delta, [, ])$ recognizing $L$. One can easily build two
  rational languages $I$ and $F$ and a set of transducers $T$ globally
  deterministic with respect to $I$ and $F$ such that $L(G,I,F) = L$
  where $G$ is the rational graph defined by $T$.  The work alphabet
  of $T$ is $\Gamma' = \Sigma \cup \{[,]\} \cup \delta$. The set of
  control states of transducer $T_a \in T$ is $\{q_0^a\} \cup
  \{q^a_{AB} \mid A,B \in \Gamma \cup \{[\} \}$, where $q_0^a$ is the
  unique initial state. Its transitions are:
  \begin{gather*}
    \forall a, b \in \Sigma, \quad q_0^a \erb{[/a} q^a_{[a} \quad
    \text{ and } \quad q_0^a \erb{b/a} q^a_{ba}
    \\
    \forall d_1 = ([,A,B,A') \in \delta, \quad q^a_{[A} \erb{[ / d_1}
    q^a_{[A'}
    \\
    \forall d_1 = (A,B,C,B'), d_2 = (B,C,D,C') \in \delta, \quad
    q^a_{BC} \erb{d_1 / d_2} q^a_{B'C'}
  \end{gather*}
  The terminal states of $T_a$ are $q_{[\bot}$ and $q_{\bot\bot}$. Now
  let $I = ([)^*$ and $F = \Sigma R^*$ where $R = \{(a,b,],b') \in
  \delta \mid a,b,b' \in \Gamma\}$. By construction and since ${C}$ is
  deterministic, $T$ is globally deterministic with respect to $I$ and
  $F$. One can easily verify that $L(G,I,F) = L$.
\end{proof}

\section{Conclusion}

\label{sect:concl}
This work is a summary of new and existing results concerning rational
graphs and their relation to context-sensitive languages. Its main
contributions are, first, to show the language equivalence between
rational graphs and synchronous rational graphs, and second to
establish a tight connection between synchronous rational graphs and
finite tiling systems. Since tiling systems accept precisely the
context-sensitive languages, this yields a new and simpler proof that
the languages of rational graphs coincide with this family.  Thanks to
this, we studied the impact of structural restrictions on the obtained
family of languages, in particular when considering finite or bounded
degree and a single initial vertex.

This approach also enables us to consider the case of deterministic
languages. We show how one can define sub-families of rational graphs
whose languages are precisely the unambiguous or deterministic
context-sensitive languages. However, due to their syntactical nature,
these results brings little new insight as to the difficult question
of the strictness of inclusions between deterministic, unambiguous and
general context-sensitive languages.

This presentation gives rise to a few interesting open questions. A
thorough study of graphs of bounded degree seems necessary, albeit
difficult. More generally, the question of knowing whether any
``tractable'' family of graphs accepting the context-sensitive
languages exists remains. We saw that synchronous graphs are not a
good option since they lose all their expressive power when only a
finite number of initial vertices are considered.  Synchronized graphs
form an interesting class, especially since their first order theory
is decidable, but it seems reasonable to believe that they require
infinite out-degree to accept all context-sensitive languages.

Another question is to compare the rational graphs with the transition
graphs of linearly bounded machines \cite{Knapik99,Payet00}. This last
point is addressed to some extent in \cite{CarayolM05}, where it is
shown that all bounded degree rational graphs are isomorphic to
transition graphs of linearly bounded machines.

\section*{Acknowledgement.}
The authors would like to thank Kamal Lodaya for his comments, and for
pointing out the interest of using tiling systems, and Didier Caucal
for his general advice and support.

\bibliographystyle{alpha}
\bibliography{legume}

\end{document}